\documentclass[mathpazo]{cicp}
\usepackage{amssymb}
\usepackage{latexsym}
\usepackage{amsthm}
\usepackage{amsbsy}
\usepackage{amsmath}
\usepackage{graphics,graphicx}
\usepackage{subfigure}
\usepackage{multirow,multicol}
\usepackage{bm}
\usepackage{bbm}
\usepackage{color}
\usepackage{float}
\usepackage{calc}
\usepackage{caption}
\usepackage{dsfont}
\usepackage{ifthen}
\usepackage{mathrsfs}
\usepackage[colorlinks,linkcolor=blue,citecolor=blue]{hyperref}
\usepackage{epstopdf}
\usepackage{epsfig}
\usepackage{algorithm}
\usepackage{algorithmic}
\usepackage{pifont}
\usepackage{overpic}
\usepackage{psfrag}
\usepackage{rotating}
\usepackage{stmaryrd}
\usepackage{verbatim}
\usepackage{setspace}
\usepackage{tikz}
\usepackage{listings} 
\usepackage{xcolor} 
\usepackage{pdfsync}
\usepackage{fancyvrb}
\usepackage{mlmath}
\usepackage{ulem}

\numberwithin{equation}{section}

\newcommand{\R}{\mathbb{R}}

\newcommand{\bxi}{\bm{\xi}}

\begin{document}
\title{A multi-scale DNN algorithm for nonlinear elliptic equations with multiple scales}

\author[X.A Li et.al~]{Xi-An Li\affil{1},Zhi-Qin John Xu\affil{1}\comma\affil{2,3}\comma\corrauth, and Lei Zhang\affil{1}\comma\affil{2,3}}
\address{\affilnum{1}\ School of Mathematical Sciences, Shanghai Jiao Tong University, Shanghai 200240, China\\
	\affilnum{2}\ Institute of Natural Sciences, Shanghai Jiao Tong University, Shanghai 200240, China\\
	\affilnum{3}\ MOE-LSC, Shanghai Jiao Tong University, Shanghai 200240, China}
\email{{\tt lixian9131@163.com, lixa0415@sjtu.edu.cn} (X.A. Li), {\tt xuzhiqin@sjtu.edu.cn} (Z.Q.J.~Xu), {\tt Lzhang2012@sjtu.edu.cn} (L.~Zhang)}

\begin{abstract}
Algorithms based on deep neural networks (DNNs) have attracted increasing attention from the scientific computing community. DNN based algorithms are easy to implement, natural for nonlinear problems, and have shown great potential to overcome the curse of dimensionality. In this work, we utilize the multi-scale DNN-based algorithm (MscaleDNN) proposed by Liu, Cai and Xu (2020) to solve multi-scale elliptic problems with possible nonlinearity, for example, the p-Laplacian problem. We improve the MscaleDNN algorithm by a smooth and localized activation function. Several numerical examples of multi-scale elliptic problems with separable or non-separable scales in low-dimensional and high-dimensional Euclidean spaces are used to demonstrate the effectiveness and accuracy of the MscaleDNN numerical scheme.
\end{abstract}


\maketitle

\section{Introduction}\label{sec:01}

In this paper, we will introduce a DNN based algorithm for the following  elliptic equation with multiple scales and possible nonlinearity 
\begin{equation}
\label{eqn:nonlinear}
\left\{
	\begin{aligned}
		-\mathrm{div} \bigg{(}a(\bm{x}, \nabla u(\bm{x}))\bigg{)} = f(\bm{x})\quad \text{in }\Omega\\
		u(\bm{x}) = g(\bm{x}) \quad \text{on }\partial\Omega
	\end{aligned}
\right.	
\end{equation}  
where $\Omega\subset \mathbb{R}^d$, $d\geq 2$, is a polygonal (polyhedral) domain (open, bounded and connected), $a(\bm{x}, \nabla u(\bm{x})): \Omega\times\mathbb{R}^d\to \mathbb{R}^d$ is the flux function, and $f:\Omega\to \mathbb{R}$ is the source term. 

Deep neural networks (DNNs) has not only achieved great successes in computer vision, natural language processing and other machine learning tasks \cite{lecun2015deep,goodfellow2016deep}, but also captured great attention in the scientific computing community due to its universal approximating power, especially in high dimensional spaces \cite{yarotsky2017error}. It has found applications in the context of numerical solution of ordinary/partial differential equations, integral-differential equations and dynamical systems \cite{qin2019data,e2018the,zhu2019physics, hutzenthaler2018overcoming,han2019uniformly,strofer2019data}.

Recent theoretical studies on DNNs have shed some light on the design of DNN-based algorithms for scientific computing tasks, in particular, for multi-scale problems. For example, the  frequency principle (F-Principle) \cite{xu_training_2018,rahaman2018spectral,xu2019frequency,e2019machine}, shows that, DNNs often fit target functions from low frequency components to high frequency ones, as opposed to the behavior of many conventional iterative numerical schemes (e.g., Gauss-Seidel method), which exhibit faster convergence for higher frequencies. To improve the convergence for high-frequency or multi-scale problems, a series of algorithms are developed to accelerate the learning of high-frequency components based on F-Principle \cite{liu2020multi,jagtap2019adaptive,cai2019phasednn,biland2019frequency}.  In particular, a multi-scale DNN algorithm(MscaleDNN) has achieved favourable performance boost for high-frequency problems \cite{liu2020multi}. The idea of the MscaleDNN to convert high-frequency contents into low-frequency ones as follows. The Fourier space is partitioned with respect to the radial direction. Since scaling input can shift the frequency distribution along the radial direction, a scaling down operation is used to scale the high-frequency components to low-frequency ones. Such radial scaling is independent of dimensionality, hence MscaleDNN is applicable for high-dimensional problems. Also, borrowing the multi-resolution concept of wavelet approximation theory using compact scaling and wavelet functions, an localized activation function (i.e., sReLU) was designed in previous work \cite{liu2020multi}, which is a product of two ReLU functions. By setting multiple scalings in a MscaleDNN, numerical results in previous study \cite{liu2020multi} show that MscaleDNN is effective for linear elliptic partial differential equations with high frequencies.

We focus our exposition on the numerical method, and therefore restrict the  flux function in \eqref{eqn:nonlinear} to the following Leray-Lions form \cite{Fractional2008growth} since it admits a natural variational form. Namely, for $(\bm{x}, \bxi)\in \Omega\times \R^d$, $\displaystyle a(\bm{x}, \bxi) = \kappa(\bm{x}) \vphi'(|\bxi|)\frac{\bxi}{|\bxi|}$, where $\vphi\in C^2$ is the so-called $N-$function (an extension for the convex function with $\vphi'(0)=0$, see \cite{Fractional2008growth} for the precise definition). For p-Laplacian problem, $\displaystyle\vphi(t) = \frac{1}{p}t^p$ , and when $p=2$ then $a(\bm{x}, \bxi)=\kappa(\bm{x}) \bxi $ becomes linear. $\kappa(\bm{x})\in L^{\infty}(\Omega)$ is symmetric, uniformly elliptic on $\Omega$, and may contain (non-separable) multiple scales. More general nonlinear flux will be treated in future work. With the above setup, the elliptic problem \eqref{eqn:nonlinear} is monotone and coercive, therefore it admits a unique solution. Those models have applications in many areas such as heterogeneous (nonlinear) materials\cite{geers2017homogenization}, non-Newtonian fluids, surgical simulation, image processing, machine learning \cite{slepev2017analysis}, \textit{etc}. 

In the last decades, much effort has been made for the numerical solution of the \eqref{eqn:nonlinear}. In particular, for p-Laplacian equation with $\kappa(\bm{x})=1$,  Some degrees of effectiveness can be achieved by mesh based methods such as finite element method (FEM) \cite{barrett1994finite, huang2007preconditioned, belenki2012optimality}, finite difference method (FDM) \cite{oberman2013finite}, discontinuous Galerkin method \cite{cockburn2016a}, and meshless methods \cite{li2018the, chaudhary2016web} \textit{etc}.. In addition to those discretization methods, iterative methods such as preconditioned steepest descent, quasi-Newton or Newton method are employed to deal with the nonlinearity. The fully nonlinear problem \eqref{eqn:nonlinear} may become singular and/or degenerate, some regularization of the nonlinear operator needs to be added to ensure the convergence of the nonlinear iteration \cite{huang2007preconditioned}. 

However, conventional methods cannot deal with the multiple scales, which is of great interest in applications for composite materials, geophysics, machine learning \textit{etc} \cite{feyel1999multiscale}. Homogenization method \cite{cioranescu2000an, tartar2009the} relies on the assumption of scale separation and periodicity. In addition, for nonlinear problems, one need to resort to a series of cell problems with the cell size going to infinity, which limits the practical utility of the method. In comparison, numerical homogenization methods, can solve linear multi-scale problems \cite{e2007heterogeneous,hou2009multiscale,owhadi2014polyharmonic} and nonlinear multi-scale problems \cite{abdulle2013analysis,chung2017a} on the coarse scale, without resolving all the fine scales. Nonetheless, the aforementioned numerical methods are easy to implement in low-dimensional space  $\mathbb{R}^d(d=1,2,3)$, however, they will encounter great difficulty in high-dimensions.

In this paper, based on the Deep Ritz method \cite{e2018the}, we proposed an improved version of the MscaleDNN algorithm  to solve elliptic problems \eqref{eqn:nonlinear} with multiple scales and/or nonlinearity. We improve the MscaleDNN by designing a new activation function due to the following intuition. The original activation function, i.e., sReLU, is localized only in the spatial domain due to the first-order discontinuity. However, the MscaleDNN requires the localization in the Fourier domain, which is equivalent to the smoothness in the spatial domain. Therefore, we design a smooth and localized activation function, which is a production of sReLU and the sine function, i.e., s2ReLU. In addition, our DNN structures also employ the residual connection technique, which was first proposed in \cite{he2016deep} for image analysis and has become very popular due to its effectiveness.  We employ this improved MscaleDNN to solve multi-scale elliptic problems, such as the multi-scale p-Laplacian equation, in various dimensions. Numerical experiments demonstrate that the algorithm is effective to obtain the oscillatory solution for multi-scale problems with or without nonlinearity, even in relative high dimensions. And the performance of s2ReLU activation function is much better than that of sReLU in the MscaleDNN framework.  

The paper is structured as follows. In Section \ref{sec:02}, we briefly introduce the framework of deep neural network approximation. Section \ref{sec:03} provides a variational formulation to solve the nonlinear multi-scale problem by MscaleDNN. In Section \ref{sec:04}, some numerical experiments are carried out to demonstrate the effectiveness of our method. Concluding remarks are made in Section \ref{sec:05}.

\section{Deep Neural Networks and ResNet architecture}\label{sec:02}

In recent years, the DNN has achieved great success in a wide range of applications, such as classification in complex systems and construction of response surfaces for high-dimensional models. Mathematically, the DNN is a nested composition of sequential linear functions and nonlinear activation functions. A standard single layer network, e.g.,  the neural unit of a DNN with a $d$-dimensional vector $\bm{x}\in \mathbb{R}^d$ as its input and a $m$-dimensional vector as its output, is in the form of
\begin{equation}\label{eq0201}
\bm{y}=\sigma(\bm{W} \bm{x}+\bm{b})
\end{equation}
where $\bm{W} = (w_{ij}) \in \mathbb{R}^{m\times d}, \bm{b}\in\mathbb{R}^{m}$ are denoted by weights and biases, respectively. $\sigma(\cdot)$ is an element-wise non-linear function, commonly known as the activation function. Various activation functions are proposed in machine learning literature, such as sigmoid, tanh, ReLU, Leaky-ReLU \textit{etc} \cite{Haykin2009comprehensive}. In DNN, the single layer \eqref{eq0201} is also denoted as the hidden layer. Its output can be transformed through new weights, biases, and activation functions in the next layer. 

Given an input datum  $\bm{x}\in\mathbb{R}^{d}$, the output of a DNN, denoted by $\bm{y}(\bm{x};\bm{\theta})$, can be written as
\begin{equation}
\bm{y}(\bm{x};\bm{\theta}) = \bm{W}^{[L]}\circ \sigma(\bm{W}^{[L-1]}\circ\sigma(\cdots\circ\sigma(\bm{W}^{[1]}\circ\sigma(\bm{W}^{[0]}\bm{x}+\bm{b}^{[0]})+\bm{b}^{[1]}))+\bm{b}^{[L-1]})+\bm{b}^{[L]}
\label{eq0202}
\end{equation}
where $\bm{W}^{[l]} \in  \mathbb{R}^{n_{l+1}\times n_l}, \bm{b}^{[l]}\in\mathbb{R}^{n_{l+1}}$ are the weights and biases of the $l$-th hidden layer, respectively. $n_0=d$, and $``\circ"$ stands for the elementary-wise operation. $\bm{\theta}$ represents the set of parameters $\bm{W}^{[L]},\cdots \bm{W}^{[1]},\bm{W}^{[0]}, \bm{b}^{[L]},\cdots \bm{b}^{[1]},\bm{b}^{[0]}$. 

Many experiments have shown that the approximating capability of the DNN will became better and more robust with increasing depth. However, the problem of gradient explosion or vanishing might occur when the depth of DNNs increases, which will have a negative effect on the performance of the DNNs. ResNet (Residual Neural Network) \cite{he2016deep} skillfully overcomes the vanishing (exploding) gradient problem in backpropagation by introducing a skip connection between input layer and output layer or some hidden layers. It makes the network  easier to train, and also improves the performance of DNN. For example, it outperforms the VGG models and obtain excellent results by using extremely deep residual nets on the ImageNet classification data set. Mathematically, a Resnet unit with $L$ layers produces a filtered version $y_N$ for the input $\bm{y}^{[0]}$ is as follows
\begin{equation*}
\bm{y}^{[\ell+1]} = \bm{y}^{[\ell]}+\sigma(\bm{W}^{[\ell+1]}\bm{y}^{[\ell]}+\bm{b}^{[\ell+1]}),~~\text{for}~~\ell=0,1,2,\cdots,N-1.
\end{equation*}
In this work, we also employ the strategy of one-step skip connection {for two consecutive layers in DNNs if they have the same number of neurons. For those consecutive layers with different neuron numbers, the skip connection step is omitted}.

\section{Unified DNN model for multi-scale problems with scale transformation}\label{sec:03}

The multi-scale elliptic problem \eqref{eqn:nonlinear} with N-function $\vphi$ admits a natural variational form \cite{ciarlet1978the}. Define the energy functional as
\begin{equation}
\label{eqn:energy}
    \mathcal{J}(v) : = \int_\Omega \kappa(\bm{x})\vphi(|\nabla v|)d\bm{x} - \int_\Omega fv d\bm{x},
\end{equation}
where $v$ is the trial function in the admissible Orlicz-Sobolev space $V:=W_g^{1,\vphi}(\Omega)$ \cite{Fractional2008growth} where the subscript ${}_g$ means that the trace on $\partial\Omega$ is $g$. The solution of \eqref{eqn:nonlinear} can be obtained by minimizing $\mathcal{J}(v)$, i.e.,
\begin{equation}\label{eqn:variational}
u = \underset{v\in V}{\arg\min}\mathcal{J}(v).
\end{equation}

Therefore, we can employ the Deep Ritz method to solve \eqref{eqn:variational}, which is an efficient approach to solve variational problems that stem from generally partial differential equations \cite{e2018the}. 

We consider an ansatz $y(\cdot;\bm{\theta})$ represented by a DNN with parameter $\theta$ as the trial function in the variational problem \eqref{eqn:variational}, where $\bm{\theta}\in\Theta$ denotes the parameters of the underlying DNN. Substituting $y(\bm{x};\bm{\theta})$ into \eqref{eqn:energy} and \eqref{eqn:variational}, we can obtain the following equation
\begin{equation}\label{minimize_integral}
u = \underset{y\in V}{\arg\min}\left[\underset{\Omega}{\int} \kappa(\bm{x})\vphi(|\nabla y(\bm{x};\bm{\theta})|) d\bm{x}-\underset{\Omega}{\int} f(\bm{x}) y(\bm{x};\bm{\theta})d\bm{x}\right].
\end{equation}
{Similar to the general strategy of searching a solution which satisfying boundary conditions of \eqref{eqn:nonlinear} in the admissible space $V$ \cite{e2018the,ciarlet1978the}}, we further approximate the integral by Monte Carlo method \cite{robert1999monte} and convert the minimization problem with respect to $y\in V$ to an equivalent one with respect to the parameters $\bm{\theta}$,  
\begin{equation}\label{mini_sample_form}
\bm{\theta}^*=\underset{\bm{\theta}\in{\Theta}}{\arg\min}\frac{1}{n_{it}}\sum_{i=1}^{n_{it}} \left[\kappa(\bm{x}_I^i)\vphi(|\nabla y(\bm{x}_I^i;\bm{\theta})|)- f(\bm{x}_I^i) y(\bm{x}_I^i;\bm{\theta})\right]~~\text{for}~~\bm{x}_I^i\in \Omega.
\end{equation}
and $y(\bm{x}, \theta) = g(\bm{x})$ for $\bm{x}\in \partial\Omega$.

Boundary conditions are indispensable constraints for numerical solution of PDEs. Analogously, imposing boundary conditions is also an important issue in DNN representation. We approximate the squared $L^2$ norm of the boundary discrepancy   $\int_{\partial\Omega} (y(\bm{x}, \theta)-g(\bm{x}))^2$ by a Monte Carlo approximation,
\begin{equation}\label{eq2bd}
\frac{1}{n_{bd}}\sum_{j=1}^{n_{bd}} \bigg{[}y\big(\bm{x}_B^j;\bm{\theta}\big)-g(\bm{x}_B^j)\bigg{]}^2   ~~\text{for}~~\bm{x}_B^j\in \partial \Omega.
\end{equation}

We define the following loss function where the boundary condition is treated as a penalty term with parameter $\beta$, 
\begin{equation}\label{eq2loss}
\mathcal{L}(\bm{\theta};{\mathcal{X}_I, \mathcal{X}_B}) = \underbrace{\frac{1}{n_{it}}\sum_{i=1}^{n_{it}} \left[\kappa(\bm{x}_I^i)\vphi(| \nabla y(\bm{x}_I^i;\bm{\theta})|)- f(\bm{x}_I^i) y(\bm{x}_I^i;\bm{\theta})\right]}_{\textup{loss\_{it}}}+
\underbrace{\frac{\beta}{n_{bd}}\sum_{j=1}^{n_{bd}} \bigg{[}y\big(\bm{x}_B^j;\bm{\theta}\big)-g(\bm{x}_B^j)\bigg{]}^2}_{\textup{loss\_{bd}}}
\end{equation}
where ${\mathcal{X}_I}=\{\bm{x}_I^i\}_{i=1}^{n_{it}}$ and ${\mathcal{X}_B}=\{\bm{x}_B^j\}_{j=1}^{n_{bd}}$ represent the training data in the iterior of $\Omega$ and on the boundary $\partial \Omega$, respectively. $\textup{loss\_{it}}$ and $\textup{loss\_{bd}}$ stand for the loss computed with the interior points and the boundary points, respectively. The penalty parameter $\beta$ controls the relative contribution of $\textup{loss\_{bd}}$ in the loss function. It increases gradually within the training process in order to better enforce the boundary condition. 
\begin{figure}
	\begin{center}
		\begin{tikzpicture}[scale=0.7]			
		\node[circle, fill=red!50,inner sep=2.25pt] (x) at  (-2.0,3){\huge$\bm{x}$};
		\node[circle, fill=blue!50,inner sep=3.5pt] (h-11) at (-0.5,0.0){};
		\node[circle, fill=blue!50,inner sep=3.5pt] (h-12) at (-0.5,1.50){};
		\node[circle, fill=blue!50,inner sep=3.5pt] (h-13) at (-0.5,3.00){};
		\node[circle, fill=blue!50,inner sep=3.5pt] (h-14) at (-0.5,4.50){};
		\node[circle, fill=blue!50,inner sep=3.5pt] (h-15) at (-0.5,6.00){};
		
		\node (connect1) at (0.35,6.00){};
		
		\draw[line width=0.8pt,color=magenta,->] (x) --node[below, rotate=-60,yshift=0.5mm,color=black]{\small$a_{1}x$} (h-11);
		\draw[line width=0.8pt,color=magenta,->] (x) --node[above,rotate=-40,yshift=-0.5mm,color=black]{\small$a_{2}x$} (h-12);
		\draw[line width=0.8pt,color=magenta,->] (x) --node[above,yshift=-1mm,color=black]{$\cdots$} (h-13);
		\draw[line width=0.8pt,color=magenta,->] (x) --node[below,rotate=35,yshift=0.5mm,color=black]{$\cdots$} (h-14);
		\draw[line width=0.8pt,color=magenta,->] (x) --node[above, xshift=0.5mm,yshift=-0.35mm,rotate=60,color=black]{\small $a_{N}x$} (h-15);
		
		\node[circle, fill=blue!50,inner sep=3.5pt] (h01) at (1.5,0.0){};
		\node[circle, fill=blue!50,inner sep=3.5pt] (h02) at (1.5,1.50){};
		\node[circle, fill=blue!50,inner sep=3.5pt] (h03) at (1.5,3.0){};
		\node[circle, fill=blue!50,inner sep=3.5pt] (h04) at (1.5,4.5){};
		\node[circle, fill=blue!50,inner sep=3.5pt] (h05) at (1.5,6.0){};
		
		\draw[line width=0.8pt,color=magenta,->] (h-11) -- (h01);
		\draw[line width=0.8pt,color=magenta,->] (h-11) -- (h02);
		\draw[line width=0.8pt,color=magenta,->] (h-11) -- (h03);
		\draw[line width=0.8pt,color=magenta,->] (h-11) -- (h04);
		\draw[line width=0.8pt,color=magenta,->] (h-11) -- (h05);
		
		\draw[line width=0.8pt,color=magenta,->] (h-12) -- (h01);
		\draw[line width=0.8pt,color=magenta,->] (h-12) -- (h02);
		\draw[line width=0.8pt,color=magenta,->] (h-12) -- (h03);
		\draw[line width=0.8pt,color=magenta,->] (h-12) -- (h04);
		\draw[line width=0.8pt,color=magenta,->] (h-12) -- (h05);
		
		\draw[line width=0.8pt,color=magenta,->] (h-13) -- (h01);
		\draw[line width=0.8pt,color=magenta,->] (h-13) -- (h02);
		\draw[line width=0.8pt,color=magenta,->] (h-13) -- (h03);
		\draw[line width=0.8pt,color=magenta,->] (h-13) -- (h04);
		\draw[line width=0.8pt,color=magenta,->] (h-13) -- (h05);
		
		\draw[line width=0.8pt,color=magenta,->] (h-14) -- (h01);
		\draw[line width=0.8pt,color=magenta,->] (h-14) -- (h02);
		\draw[line width=0.8pt,color=magenta,->] (h-14) -- (h03);
		\draw[line width=0.8pt,color=magenta,->] (h-14) -- (h04);
		\draw[line width=0.8pt,color=magenta,->] (h-14) -- (h05);
		
		\draw[line width=0.8pt,color=magenta,->] (h-15) -- (h01);
		\draw[line width=0.8pt,color=magenta,->] (h-15) -- (h02);
		\draw[line width=0.8pt,color=magenta,->] (h-15) -- (h03);
		\draw[line width=0.8pt,color=magenta,->] (h-15) -- (h04);
		\draw[line width=0.8pt,color=magenta,->] (h-15) -- (h05);
		
		\node[circle, fill=cyan!60,inner sep=1.0pt] (otimes) at (2.75, 3) {$\bm{\oplus}$};					
		\draw[line width=0.8pt,color=magenta,->] (h01) -- (otimes);		
		\draw[line width=0.8pt,color=magenta,->] (h02) -- (otimes);		
		\draw[line width=0.8pt,color=magenta,->] (h03) -- (otimes);		
		\draw[line width=0.8pt,color=magenta,->] (h04) -- (otimes);		
		\draw[line width=0.8pt,color=magenta,->] (h05) -- (otimes);
		\draw[line width=0.8pt,color=cyan,->] (h-15)..controls(0.5,7.0) and (2.25,8.5).. (otimes);
		\draw[line width=0.8pt,color=cyan,->] (h-14)..controls(0.5,7.0) and (2.25,8.5).. (otimes);
		\draw[line width=0.8pt,color=cyan,->] (h-13)..controls(0.5,7.0) and (2.25,8.5).. (otimes);
		\draw[line width=0.8pt,color=cyan,->] (h-12)..controls(0.5,7.0) and (2.25,8.5).. (otimes);
		\draw[line width=0.8pt,color=cyan,->] (h-11)..controls(0.5,7.0) and (2.25,8.5).. (otimes);
		
		\node[rectangle, fill=blue!50,inner sep=3.5pt,rotate=90, text width = 1.5cm, minimum height = 0.5cm, align = center] (h11) at (3.75,3.0){$\cdots\cdots$};
		\draw[line width=0.8pt,color=magenta,->] (otimes) -- (h11);
		
		\node[circle, fill=blue!50,inner sep=3.5pt] (h21) at (4.95,0.0){};
		\node[circle, fill=blue!50,inner sep=3.5pt] (h22) at (4.95,1.50){};
		\node[circle, fill=blue!50,inner sep=3.5pt] (h23) at (4.95,3.00){};
		\node[circle, fill=blue!50,inner sep=3.5pt] (h24) at (4.95,4.5){};
		\node[circle, fill=blue!50,inner sep=3.5pt] (h25) at (4.95,6.0){};
		\node[circle, fill=green!50,inner sep=0.5pt,text width=1.1cm] (u) at (7.0,3){$y(\bm{x};\bm{\theta})$};
		
		\draw[line width=0.8pt,color=magenta,->] (h11) -- (h21);
		\draw[line width=0.8pt,color=magenta,->] (h11) -- (h22);
		\draw[line width=0.8pt,color=magenta,->] (h11) -- (h23);
		\draw[line width=0.8pt,color=magenta,->] (h11) -- (h24);
		\draw[line width=0.8pt,color=magenta,->] (h11) -- (h25);
		
		\draw[line width=0.8pt,color=magenta,->] (h21) -- (u);
		\draw[line width=0.8pt,color=magenta,->] (h22) -- (u);
		\draw[line width=0.8pt,color=magenta,->] (h23) -- (u);
		\draw[line width=0.8pt,color=magenta,->] (h24) -- (u);
		\draw[line width=0.8pt,color=magenta,->] (h25) -- (u);
		
		\node (DNN) at (2.5,-0.75){DNN with ResNet};
		\node[circle, fill=green!60,inner sep=0.1pt] (gradU) at (8.75, 5) {\small$\nabla y(\bm{x};\bm{\theta})$};					
		
		\node[circle, fill=green!60,inner sep=1.0pt] (Ax) at (7.5, 7.5) {$\kappa(\bm{x})$};
		\node[circle, fill=green!60,inner sep=1.25pt] (fx) at (10, 8) {$f(\bm{x})$};
		\draw[line width=0.8pt,,color=black, ->] (x)..controls(-1.5,9) and (4,8.25)..(Ax);
		\draw[line width=0.8pt,,color=black, ->] (x)..controls(-1.5,9) and (4,8.5)..(fx);	
		
		\node[rectangle, fill=yellow!60, inner sep=3pt](loss-bd) at (12,1.25) {$|y(\bm{x};\bm{\theta})-g(\bm{x})|^2~$ on $\bm{x}\in\partial \Omega$};
		\node[rectangle, fill=yellow!0, inner sep=2pt](loss-bd11) at (12.25,0.3) {$loss\_{bd}$};		
		\node[rectangle, fill=yellow!60, inner sep=2.5pt](loss-it1) at (15.85,5.0) {$\displaystyle\int \kappa(\bm{x})\vphi(|\nabla y(\bm{x};\bm{\theta})|)-f(\bm{x})y(\bm{x};\bm{\theta})d\bm{x}$};
		\node[rectangle, fill=yellow!0, inner sep=2pt](loss-it11) at (15.5,6.25) {$loss\_{it}$};

		\draw[line width=0.8pt,color=black, ->] (u) -- (gradU);
		\draw[line width=0.8pt,color=black, ->] (u) -- (loss-bd);		
		
		\draw[line width=0.8pt,color=black, ->] (Ax)..controls(8,7.25) and (10,5)..(loss-it1);
		\draw[line width=0.8pt,color=black, ->] (fx)..controls(10,6.5) and (10.5,5)..(loss-it1);
		\draw[line width=0.8pt,color=black,->] (gradU) -- (loss-it1);
		\draw[line width=0.8pt,color=black, ->] (u) ..controls(9,3.5) and (10,5).. (loss-it1);	
		
		\node[rectangle, fill=yellow!0, inner sep=0pt](loss-it-out) at (15.15,4.35) {};
		\node[rectangle, fill=yellow!0, inner sep=1pt](loss-bd-out) at (15.25,1.25){};
		\node[circle, fill=cyan!60, inner sep=2pt](loss) at (17.5,2.25){loss};
		\node[circle, fill=cyan!60, inner sep=2pt](theta) at (17.5,-0.5){$\bm{\theta}^*$};	
		
		\draw[line width=0.8pt,,color=black, ->] (loss-it-out) ..controls(15.65,3) and (16.25,2.5).. (loss);
		\draw[line width=0.8pt,,color=black, ->] (loss-bd-out) ..controls(15.5,2.25) and (16.25,2.5).. (loss);
		
		\draw[line width=0.8pt,color=black,->] (loss) -- node[right]{minimize} (theta);		
		\end{tikzpicture}
	\end{center}
	\caption{Schematic of a DNN for solving the nonlinear multi-scale problem}
\end{figure}
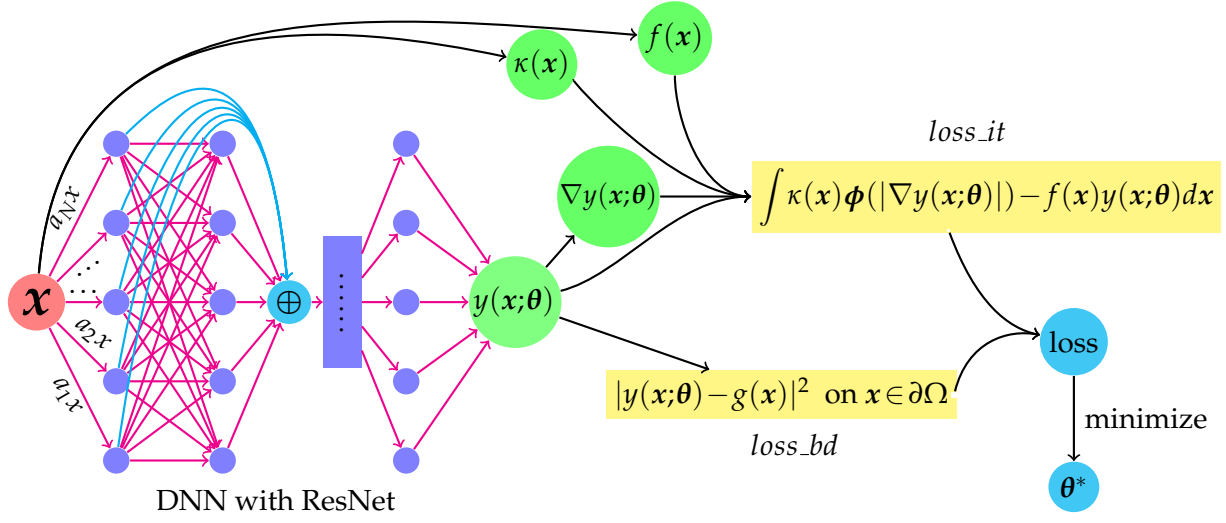
Our goal is to find a set of parameters $\bm{\theta}$ such that the approximate function $y(\bm{x};\bm{\theta})$ minimizes the loss function $\mathcal{L}(\bm{\theta})$. If $\mathcal{L}(\bm{\theta})$ is small enough, then $y(\bm{x};\bm{\theta})$ will be a "good" approximate solution of \eqref{eqn:nonlinear}, i.e.,
\begin{equation*}
\bm{\theta}^* = \arg\min \mathcal{L}(\bm{\theta})~~\Longleftrightarrow~~ u(\bm{x})\approx y(\bm{x};\bm{\theta}^*).
\end{equation*}
In order to obtain $\bm{\theta}^*$, one can update the parameter $\bm{\theta}$ by the gradient descent method over the all training examples at each iteration. The objective function decreases along a descent direction $w^k$ after an iteration, i.e., $\mathcal{L}(\bm{\theta}^{k+1})<\mathcal{L}(\bm{\theta}^k)$, where $\bm{\theta}^{k+1} = \bm{\theta}^k + \eta w^k$ with some properly chosen learning rate or step size $\eta$. Since DNNs are highly non-convex,  $\bm{\theta}^*$ may only converge to a local minimum. Stochastic gradient descent (SGD), as the common optimization technique of deep learning, has been proven to be very effective in practice (can avoid the problem of local minimum) and is a fundamental building block of nearly all deep learning models. In the implementation, SGD method chooses a "mini-batch" of data {$\mathcal{X}^{k}_m$} (a subset of interior points and boundary points in our case) at each step. In this context, the SGD is given by:
\begin{equation*}
\bm{\theta}^{k+1}=\bm{\theta}^{k}-\eta^k\nabla_{\bm{\theta}}\mathcal{L}(\bm{\theta}^{k};{\mathcal{X}^{k}_m}),
\end{equation*}
where the ``learning rate'' $\eta^k$ decreases with $k$ increasing.
\begin{remark}
   	The $\bm{\theta}$ consist of weights and biases in our model are initialized by using the normal distribution  $\mathcal{N}\left(0,\left(\frac{2}{n_{in}+n_{out}}\right)^2\right)$,
	where $n_{in}$ and $n_{out}$ are the input and output dimensions of the corresponding layer, respectively. 
\end{remark}

\subsection{MscaleDNN}
A conventional DNN model can achieve a satisfactory solution for PDE problems when the coefficient $\kappa(\bm{{x}})$ is homogeneous {(e.g., smooth or possessing few scales)} \cite{wang2020a, he2020a, berg2018a, sirignano2018dgm}. However, it is difficult to solve PDEs \eqref{eqn:nonlinear} with multi-scale $\kappa(\bm{{x}})$ due to the complex interaction of nonlinearity and multiple scales. The MscaleDNN architecture has been proposed to approximate the solution with high frequency and multiple scales by converting original data to a low frequency space \cite{liu2020multi}, as described in the following:
\begin{itemize}
	\item Divide the neurons in the first hidden-layer into $N$ groups, and generate the scale vector 
	\begin{equation}
	  \label{Separate_neurons}
	K = (a_1,\cdots,a_1, a_2,\cdots,a_2,\cdots,a_N,\cdots,a_N)^T, a_i\geqslant 1.
	\end{equation}
	{Note that the scale parameters $a_i$'s are hyper-parameters and can be set by several methods. Numerical results in previous work \cite{cai2019multi,liu2020multi} show that the effectiveness of the MscaleDNN is not sensitive to the selection of scale parameters if it covers the scales of the target function.}
	
	\item Convert the input data $\bm{x}$ to $\tilde{\bm{x}} = K\odot \bm{x}$, then feed $\tilde{\bm{x}}$ to the first hidden-layer of DNN, where $\odot$ is the Hadamard product. From the viewpoint of Fourier transformation and decomposition, the Fourier transform $\hat{f}(k)$ of a given band-limited function $f(\bm{x})$ has a compact support $B(K_{max})=\left\{k\in\mathbb{R}^d,|k|\leqslant K_{max}\right\}$ which can be partitioned as the union of $M$ concentric annulus with uniform or nonuniform width, e.g., 
    	\begin{equation*}
    	    P_i=\left\{k\in\mathbb{R}^d,(i-1)K_0\leqslant|k|\leqslant iK_0\right\},K_0=K_{max}/M,i\leqslant i\leqslant M,
    	\end{equation*}
    	then $\hat{f}(k)$ can be expressed as follows 
    	\begin{equation*}
    	    \hat{f}(k) = \sum_i^M \hat{f}_i(k),~~\textup{with}~~\mathrm{supp} \hat{f}_i(k)\subset P_i
    	\end{equation*}
    	By the down-scaling operation which shift the high frequency region $P_i$ into a low frequency ones, the scale transfrom reads, 
	   \begin{equation*}
	       \hat{f}_i^{(scale)}(k)=\hat{f}(a_i k), a_i>1
	   \end{equation*}
	   and
	   \begin{equation*}
	       f^{(scale)}_i(\bm{x})=f_i\left(\frac{1}{a_i}\bm{x}\right)
	   \end{equation*}
	   or
	   \begin{equation*}
	       f_i(\bm{x})=f_{i}^{(scale)}(a_i\bm{x}).
	   \end{equation*}
	  Then, instead of finding a function $\hat{f}_i(k)$ in the support set of $P_i$, the transformed function  $\hat{f}_i^{(scale)}(k)$ will be explored  in 
	  \begin{equation*}
	      \mathrm{supp} \hat{f}_i^{(scale)}(k) \subset \left\{k\in\mathbb{R}^d, \frac{(i-1)K_0}{\alpha_i}\leqslant |k|\leqslant \frac{iK_0}{\alpha_i}\right\}.
	  \end{equation*}
	  When $\displaystyle \frac{iK_0}{\alpha_i}$ is small, a DNN $\ell_i(\theta)$ can be used to learn $f^{(scale)}_i(x)$  quickly, which further means that $\ell_i(\theta)$ can approximate $f_i(\bm{x})$ immediately, i.e.,
	  \begin{equation*}
	     f_i(\bm{x}) \approx \ell_i(a_i\theta).
	  \end{equation*}
	   Finally, $f(\bm{x})$ can be expressed by 
	   \begin{equation}
	      \label{convert}
	       f(\bm{x})\approx\sum_i^M \ell_i(a_i\theta).
	   \end{equation}
	   In general, \eqref{convert} suggests an ansatz to approximate the solution more quickly with DNN representations, hence, converting the original data $\bm{x}$ into $\tilde{\bm{x}}$ is a nice trick when dealing with multi-scale problems.
	\item Output the result of DNN.
\end{itemize}

\subsection{Activation Function}
The role of activation functions is to decide whether particular neuron should fire or not. When the activation function is absent, the neural network will simply be a linear transformation involving weights and biases, which in turn becomes a linear regression model. Although linear model is simple to solve, but its expressive power for complex problems is limited. A nonlinear activation function performs the nonlinear transformation of the input data, making it capable to learn and perform
more complex tasks. Thus, choosing the right activation function is essential for the efficiency and accuracy of the DNN. The significance of activation functions for different models have been investigated in, e.g.,  \cite{qian2018adaptive,jagtap2020adaptive} \textit{etc}.

In the previous work \cite{liu2020multi}, the activation function $\text{sReLU}(x) = \text{ReLU}(x)\times \text{ReLU}(1-x)$ smoother than $\text{ReLU}(x) = \max\{0,x\}$ is used in the MscaleDNN algorithm to solve PDEs. It is localized in the spatial domain due to the first-order discontinuity, but it lacks of adequate smoothness. To improve the efficiency of scale separation, we propose a smoother activation function which is a production of sReLU and the sine function, and is referred to sin-sReLU,  
	\begin{equation}
	\text{sin-sReLU}(x) = \text{sin}(2\pi x)*\text{ReLU}(x)*\text{ReLU}(1-x). 
	\end{equation}
	For convenience, we abbreviate it by s2ReLU here and thereafter.
	\begin{figure}[H]
		\centering
		\subfigure[]{\label{actfunc.a}
			\includegraphics[scale=0.425]{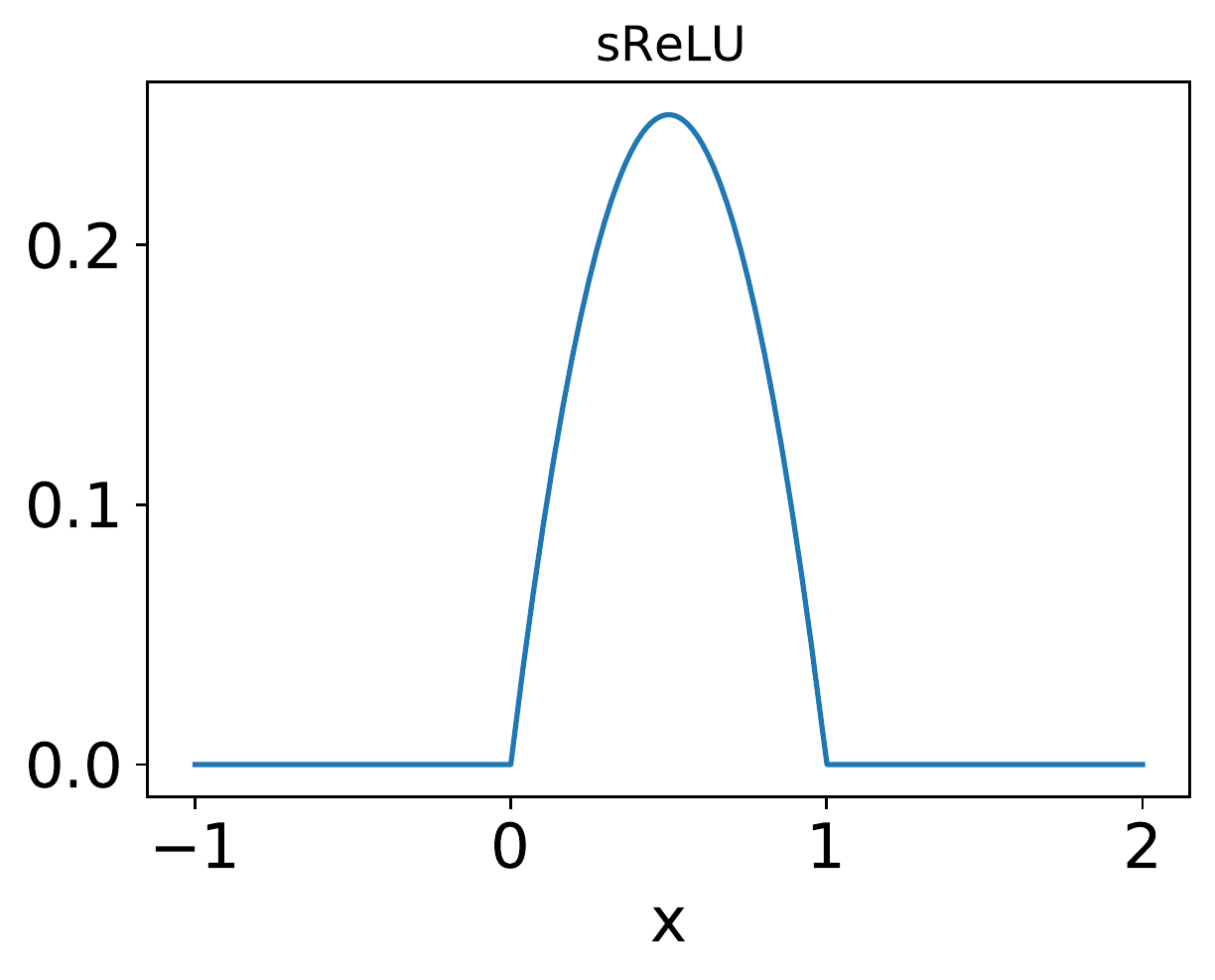}	
		}
		\subfigure[]{\label{actfunc.b}
			\includegraphics[scale=0.425]{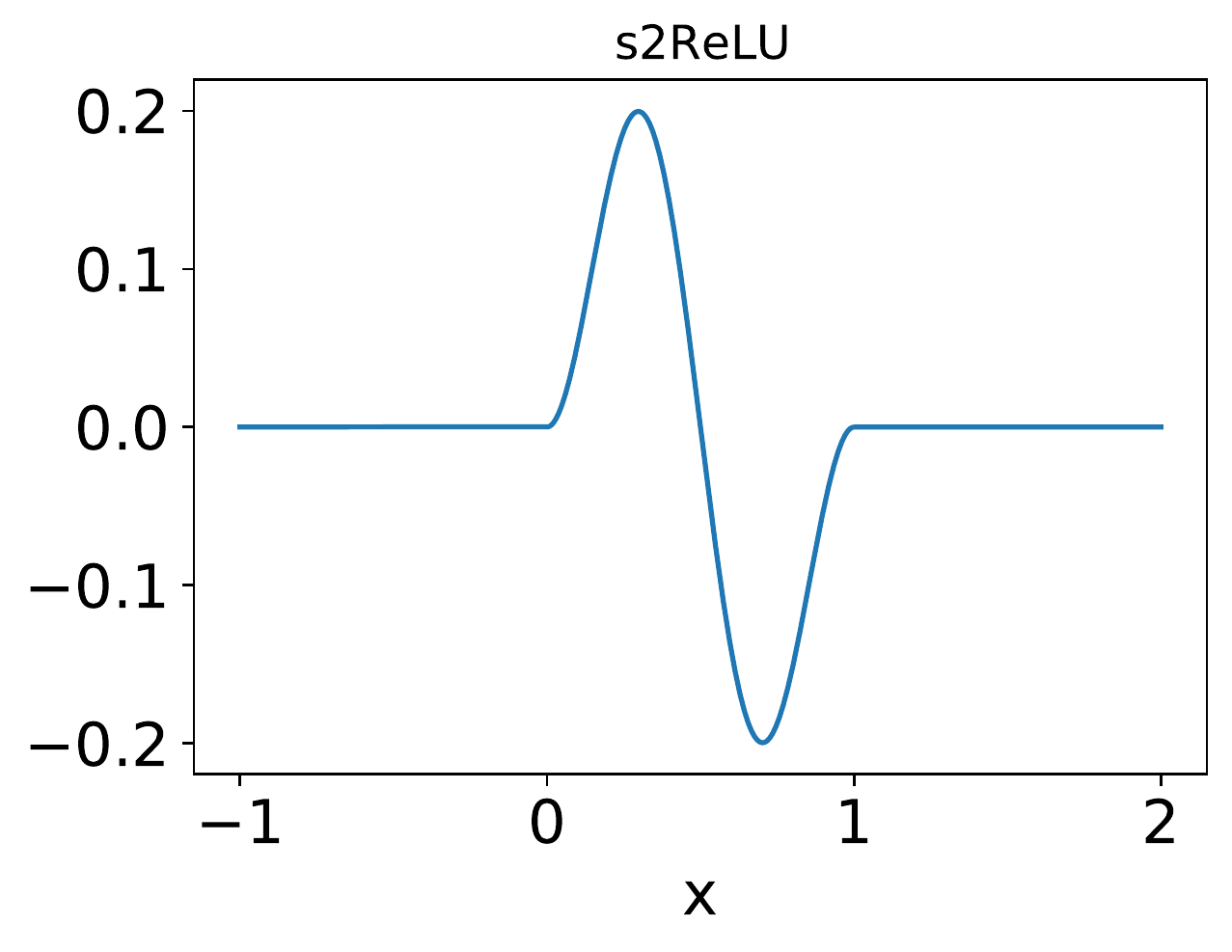}
		}
		
		\subfigure[]{\label{actfunc.c}
			\includegraphics[scale=0.425]{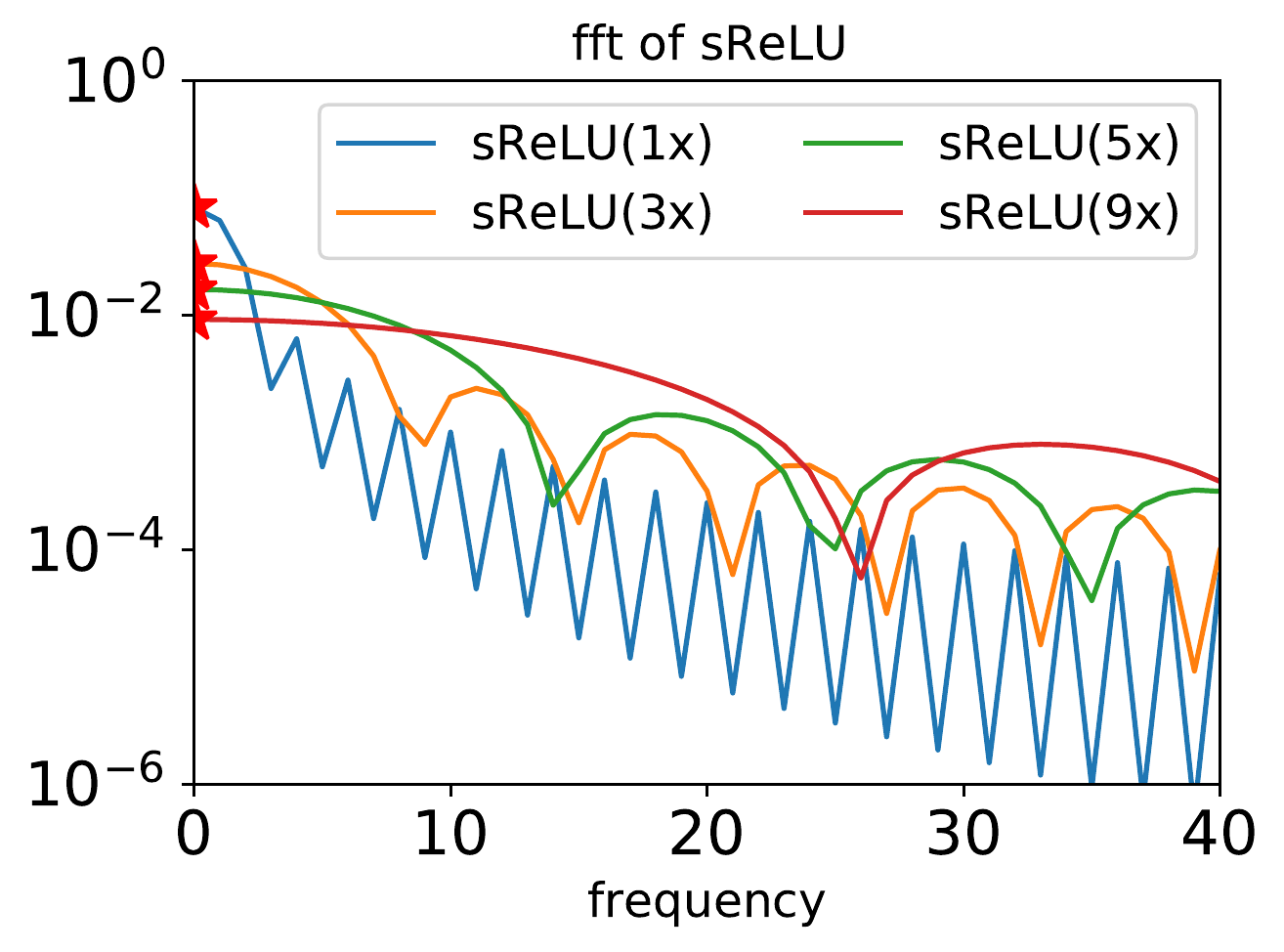}		
		}
		\subfigure[]{\label{actfunc.d}
			\includegraphics[scale=0.425]{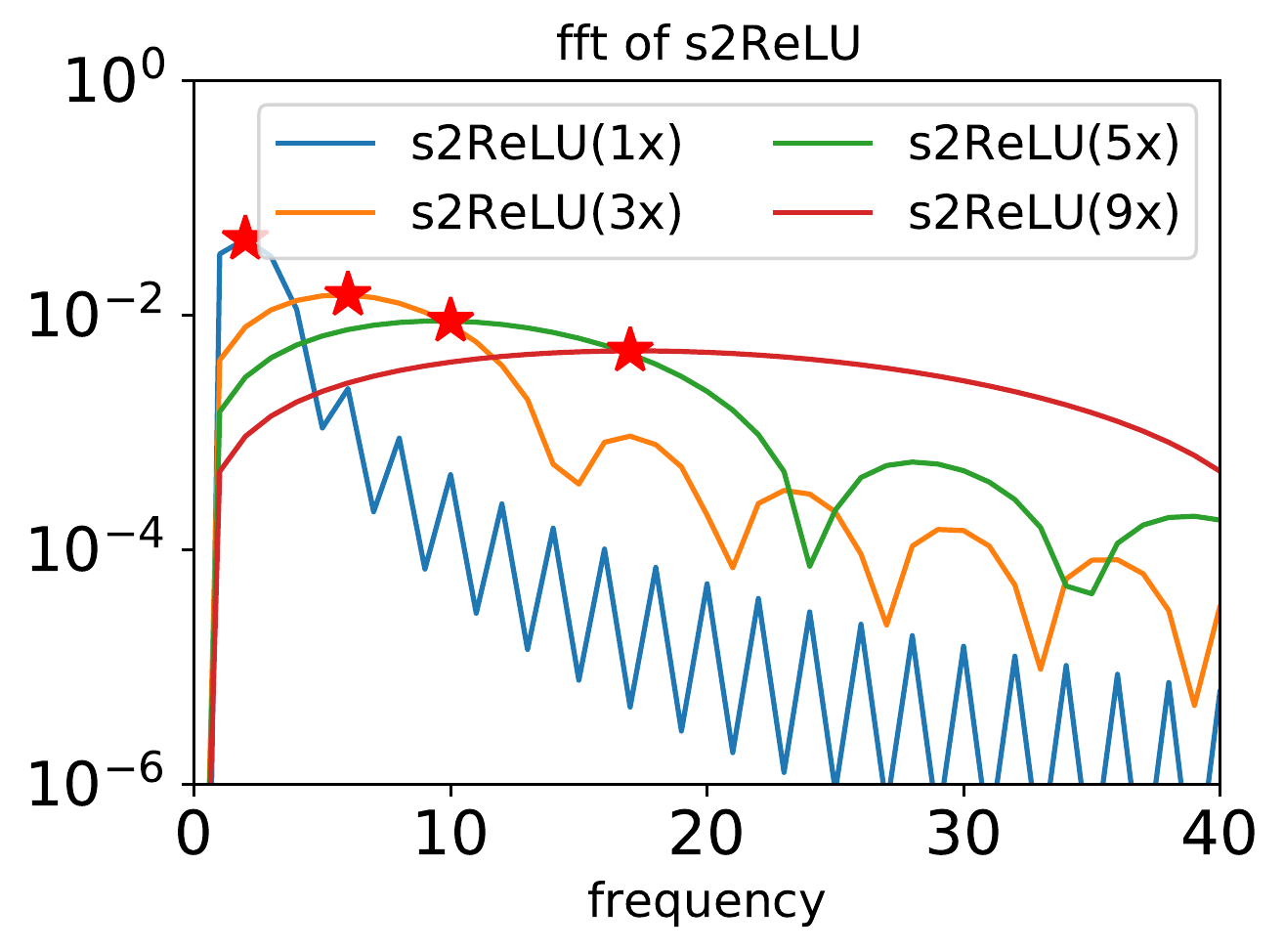}	
		}
		\caption{sReLU function (left) and s2ReLU function (right)} in the spatial (upper) and frequency (lower, the peak of each line is indicated by a star.) domains. \label{actfunc}
	\end{figure}
	
	{As shown in  in Figs. \ref{actfunc.a} and \ref{actfunc.b}, sReLU and s2ReLU are localized in the spatial domain. For sReLU, since its first-order derivative is discontinuous, its Fourier transform decays slowly. As shown in Fig. \ref{actfunc.c}, sReLU functions with different scales overlap with each other. However, since s2ReLU is a smooth function, it decays  faster and has better localization property in the frequency domain, compared with sReLU, as shown in Fig. \ref{actfunc.d}. The localization in the frequency domain, leads to the fact, that the peak-amplitude frequencies of different scaled s2ReLU functions (stars in Fig. \ref{actfunc.d}) are separated and increase as the scales increase. Therefore, s2ReLU potentially could be more efficient to represent multi-scale functions.  In the numerical experiments, we will show that s2ReLU has more superior performance compared to sReLU with MscaleDNN or ReLU with conventional DNN.}

\section{Numerical experiments}\label{sec:04}
In this section, several test problems are presented to illustrate the effectiveness of our method to solve multi-scale nonlinear problems. In our numerical experiments, all training and testing data are generated  with a uniform distribution over corresponding domains, and all networks are trained by the Adam optimizer. In addition, the initial learning rate is set as $2\times 10^{-4}$ with a decay rate $5 \times 10^{-5}$ for each training step. For the first hidden-layer in MscaleDNN, we divide the neurons into $N=100$ groups to generate the scale vector ${K=\{1, 1, 2, 3,\cdots, 99\}}$ as in  \eqref{Separate_neurons}. Here, we provide two criteria to evaluate our model:
\begin{equation*}
MSE = \frac{1}{n_s}||\tilde{u}-u^*||^{2}_{2}, ~~\text{and}~~ REL = \frac{1}{n_s}\frac{||\tilde{u}-u^*||^{2}_{2}}{||u^*||^{2}_{2}}
\end{equation*}
where $\tilde{u}$ and $u^*$ are approximate solutions of deep neural network and the exact solution (or the reference solution computed on a very fine mesh), respectively, at $n_s$ testing sample points. To evaluate the effectiveness, we test our model for every 1000 epochs in the training process. In our work, the penalty parameter $\beta$ is set as follows,
\begin{equation}
    \beta=\left\{
     \begin{aligned}
        \beta_0, \quad &\textup{if}~~i<M_{\max}*0.1\\
        10\beta_0,\quad &\textup{if}~~M_{\max}*0.1<=i<M_{max}*0.2\\
        50\beta_0, \quad&\textup{if}~~ M_{\max}*0.2<=i<M_{max}*0.25\\
        100\beta_0, \quad&\textup{if}~~ M_{\max}*0.25<=i<M_{max}*0.5\\
        200\beta_0, \quad&\textup{if}~~ M_{\max}*0.5<=i<M_{max}*0.75\\
        500\beta_0, \quad&\textup{otherwise},
     \end{aligned}
     \right.
\end{equation}
where the $\beta_0=1000$ in all our tests and $M_{\max}$ represents the total number of epoches. We perform neural network training and testing in TensorFlow (version 1.4.0) on a work station with 256GB RAM and a single NVIDIA GeForce GTX 2080Ti 12GB.

For the numerical examples, we use the following p-Laplacian problem as an prototypical example of the more general form \eqref{eqn:nonlinear},
\begin{equation}\label{eqn:pLaplacian}
\begin{cases}
-\text{div}\bigg{(}\kappa(\bm{x})|\nabla u(\bm{x}) |^{p-2} \nabla u(\bm{x})\bigg{)} = f(\bm{x}),~\bm{x}\in \Omega,\\
~~~~~~~~~~~~~u(\bm{x}) = g(\bm{x}),~~~~~~~~~~~~~~~~~~~~~~\bm{x}\in \partial\Omega,
\end{cases}
\end{equation}
then the energy in \eqref{eqn:energy} can be rewritten as
\begin{equation*}
	\mathcal{J}(v) : = \frac{1}{p}\int_\Omega \kappa(\bm{x})|\nabla v|^pd\bm{x} - \int_\Omega fv d\bm{x},~v\in V.
\end{equation*}
with $V:=W_g^{1,p}$, namely, the Sobolev space $W^{1,p}$ with trace $g$ on $\partial\Omega$.

\subsection{One dimensional examples}
We consider the following one-dimensional highly oscillatory elliptic problem in $\Omega=[0,1]$.
\begin{equation}\label{1dexample01}
\begin{cases}
\displaystyle -\frac{d}{dx}\left(\kappa(x)\bigg{|}\frac{d}{dx}u_{\epsilon}(x)\bigg{|}^{p-2}\frac{d}{dx}u_{\epsilon}(x)\right)=f(x),\\
u_{\epsilon}(0)=u_{\epsilon}(1)=0.
\end{cases}
\end{equation}

For equation \eqref{1dexample01}, we consider $\epsilon=0.1$ and $\epsilon=0.01$. We use the MscaleDNN models with activation functions sReLU and s2ReLU to solve this problem, respectively. In addition, a DNN model with ReLU is used as a baseline for comparison. 
At each training step, we uniformly sample $n_{it}=3000$ interior points in $\Omega$ and $n_{bd}=500$ boundary points on $\partial\Omega$ as the training dataset, and uniformly sample $n_s=1000$ points in $\Omega$ as the testing dataset.

\begin{example}\label{e0101}
	We consider the case of $p=2$ for the linear diffusion problem with highly oscillatory coefficients \eqref{1dexample01}. $f\equiv 1$ and,
	\begin{equation}\label{example02}
	\kappa(x)=\left(2+\cos\left(2\pi\frac{x}{\epsilon}\right)\right)^{-1}
	\end{equation}
	with a small parameter $\epsilon>0$ such that $\epsilon^{-1}\in\mathbb{N}^+$. In one-dimensional setting, the corresponding unique solution is given by
	\begin{equation}\label{example03}
	u_\epsilon(x) = x-x^2+\epsilon\left(\frac{1}{4\pi}\sin\left(2\pi\frac{x}{\epsilon}\right)-\frac{1}{2\pi}x\sin\left(2\pi\frac{x}{\epsilon}\right)-\frac{\epsilon}{4\pi^2}\cos\left(2\pi\frac{x}{\epsilon}\right)+\frac{\epsilon}{4\pi^2}\right).
	\end{equation}
\end{example}

\begin{figure}[H]
	\centering
	\subfigure[$\epsilon=0.1$]{
		\label{u}
		\includegraphics[scale=0.325]{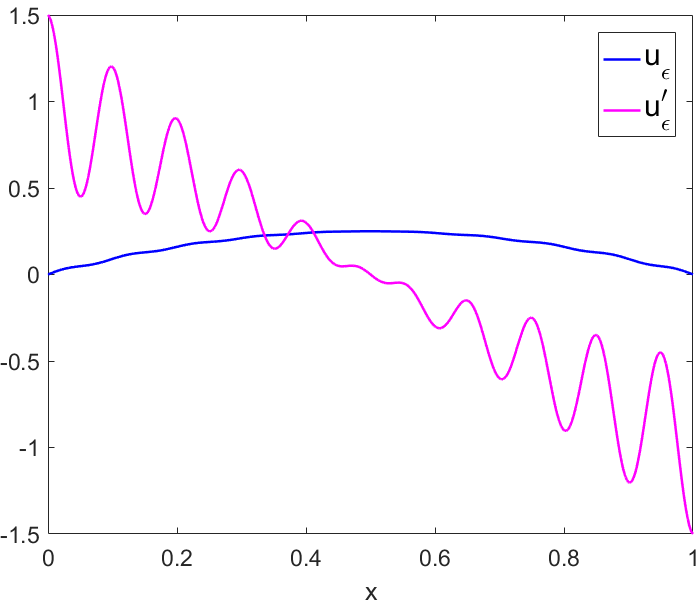}
	}
	\subfigure[$\epsilon=0.01$]{
		\label{diffu}
		\includegraphics[scale=0.325]{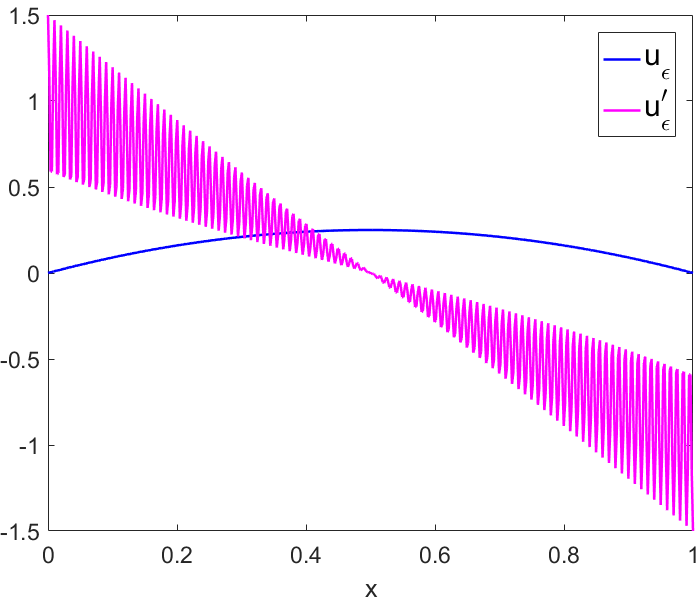}
	}
	\caption{The graphs for the original function and the derivative of $u_{\epsilon}$}
	\label{udiff}
\end{figure}

{ Since the oscillation amplitude is small, to show the highly oscillation, we display the first-order derivative of the target functions  for $\epsilon=0.1$ and $\epsilon=0.01$ in Fig. \ref{udiff}, respectively.}

\begin{figure}[H]
	\centering
	\subfigure[solution]{
		\label{p2e01a}
		\includegraphics[scale=0.365]{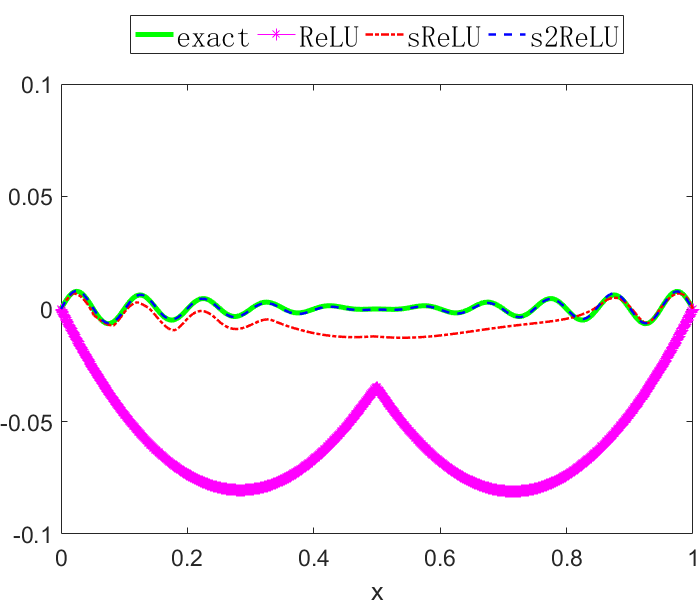}
	}
	\subfigure[MSE and REL]{
		\label{p2e01b}
		\includegraphics[scale=0.375]{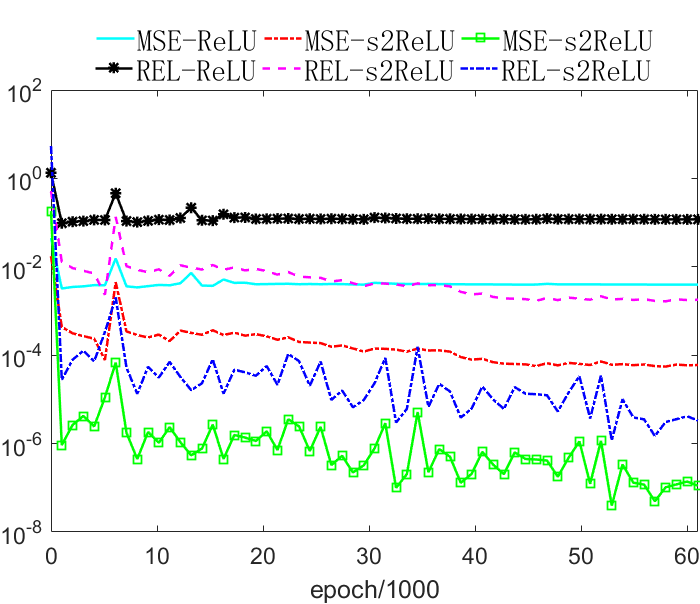}
	}
	\caption{Testing results for $\epsilon=0.1$ when $p=2$. The network size is (300, 200, 150, 150, 100, 50, 50).}
	\label{p2e01}
\end{figure}

\begin{figure}[H]
	\centering
	\subfigure[solution]{
		\label{p2e001a}
		\includegraphics[scale=0.375]{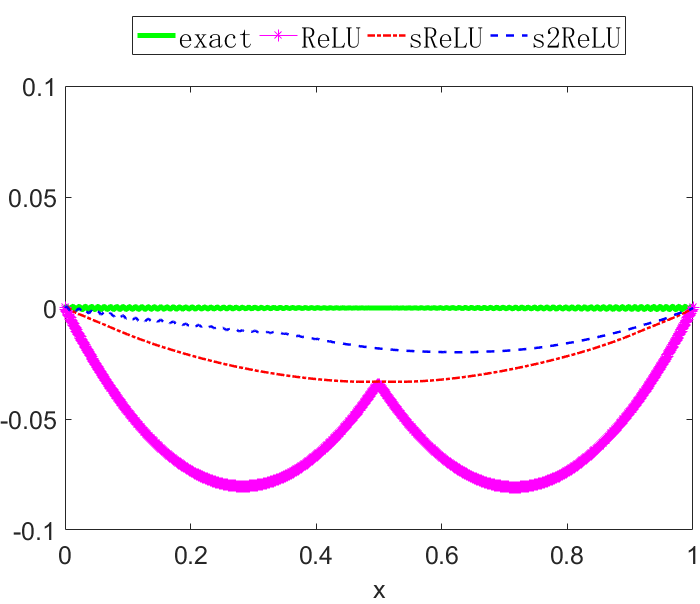}
	}
	\subfigure[MSE and REL]{
		\label{p2e001b}
		\includegraphics[scale=0.375]{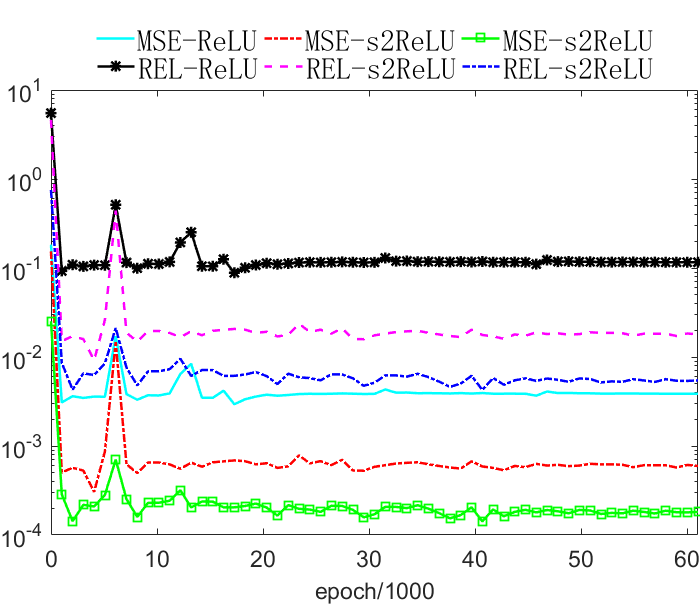}
	}
	\caption{Testing results for $\epsilon=0.01$ when $p=2$. The network size is (500, 400, 300, 300, 200, 100, 100).}
	\label{p2e001}
\end{figure}

Although the p-Laplacian equation is reduced to a linear one, the problem is still difficult to deal with by DNN due to the highly oscillatory coefficients with small $\epsilon$ \cite{xu2019frequency}. {Since the solution is a smooth $O(1)$ function with a oscillating perturbation of $O(\epsilon)$ for our one-dimensional problems, in the following, we then only illustrate the $O(\epsilon)$ parts of the solutions by subtracting $u(x)-x(1-x)$}. For $\epsilon=0.1$ as shown in Fig. \ref{p2e01a}, the solution of the MscaleDNN with activation function s2ReLU overlaps with the exact solution, while the one with sReLU deviates from the exact solution at the central part and the one with ReLU is completely different from the exact solution. As shown in Fig. \ref{p2e01b}, both the error and the relative error consistently show that MscaleDNN with s2ReLU can resolve the solution pretty well. For the case of $\epsilon=0.01$ in Figs. \ref{p2e001a}, the s2ReLU solution and the sReLU  solution both deviate from the exact solution at the central part of $(0,1)$, but the s2ReLU solution still outperforms that of sReLU. The error curves in Fig. \ref{p2e001b}  enhance this conclusion. Figs. \ref{p2e01} and \ref{p2e001} clearly reveal that the performances of MscaleDNN model with s2ReLU and sReLU are superior to that of general DNN model with ReLU.

When $p$ increases, the nonlinearity of the p-Laplacian problem \eqref{eqn:nonlinear} becomes more and more significant and has complex interaction with the highly oscillatory coefficients, hence the solution becomes increasingly more difficult. In following examples, we further consider the 1d example \eqref{1dexample01} with $p=5$, respectively. 

\begin{example}\label{1dE05}
	For $p=5$, this p-Laplacian equation is a highly oscillatory diffusion problem. The exact solution $u_{\epsilon}(x)$ and $\kappa(x)$ are the same as that of Example \ref{e0101}. The force side $f$ is given by 
	\begin{equation}\label{1dexample0404}
	\displaystyle f(x) =\frac{-|2x-1|^3\left[2+\cos(2\pi \frac{x}{\epsilon})\right]^2\left[3\pi(2x-1)\sin\left(2\pi\frac{x}{\epsilon}\right)-4\epsilon\cos\left(2\pi\frac{x}{\epsilon}\right)-8\epsilon\right]}{8\epsilon},
	\end{equation}
	where  $\epsilon>0$ and $\epsilon^{-1}\in\mathbb{N}^+$.
\end{example}

\begin{figure}[H]
	\centering
	\subfigure[solution]{
		\label{p5e01a}
		\includegraphics[scale=0.365]{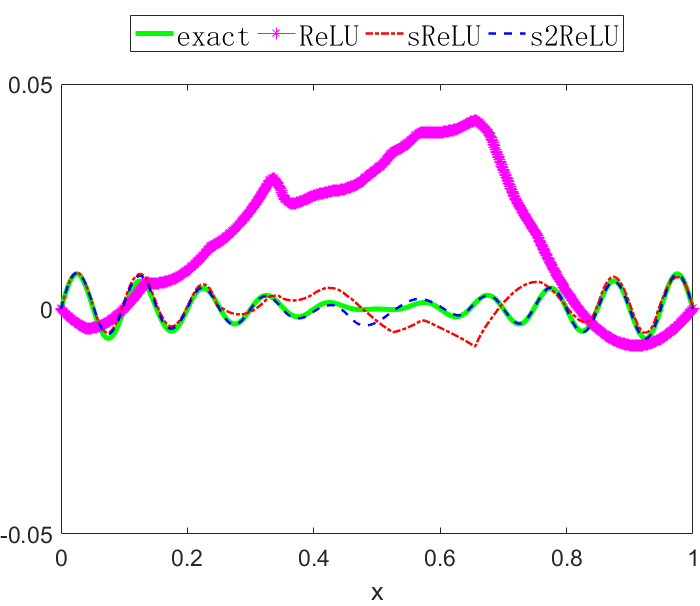}
	}
	\subfigure[MSE and REL]{
		\label{p5e01b}
		\includegraphics[scale=0.375]{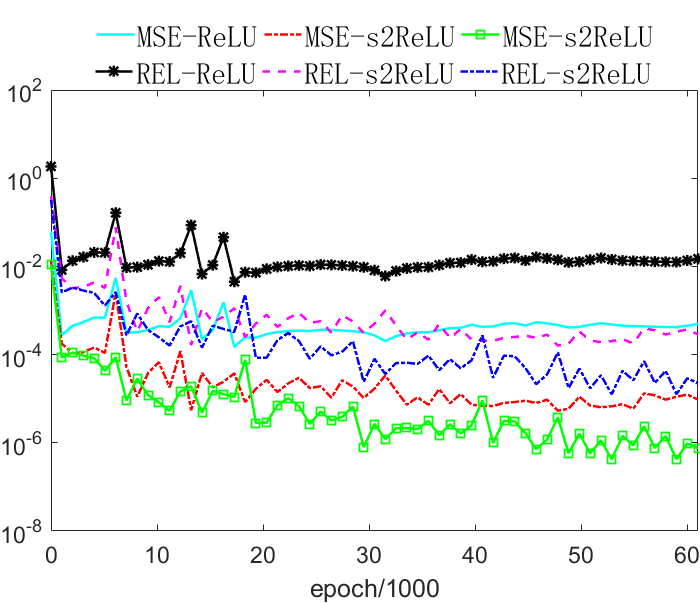}
	}
	\caption{Testing results for $\epsilon=0.1$ when $p=5$. The network size is (300, 200, 150, 150, 100, 50, 50).}
	\label{p5e01}
\end{figure}

\begin{figure}[H]
	\centering
	\subfigure[solution]{
		\label{p5e001a}
		\includegraphics[scale=0.365]{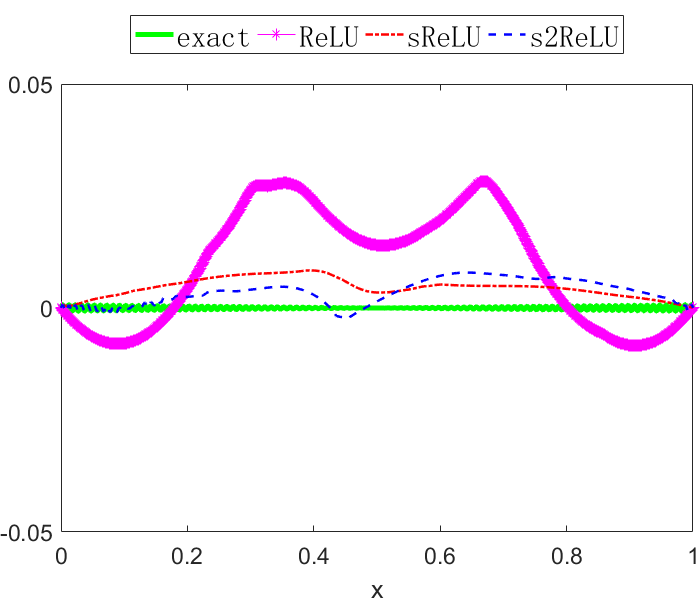}
	}
	\subfigure[MSE and REL]{
		\label{p5e001b}
		\includegraphics[scale=0.375]{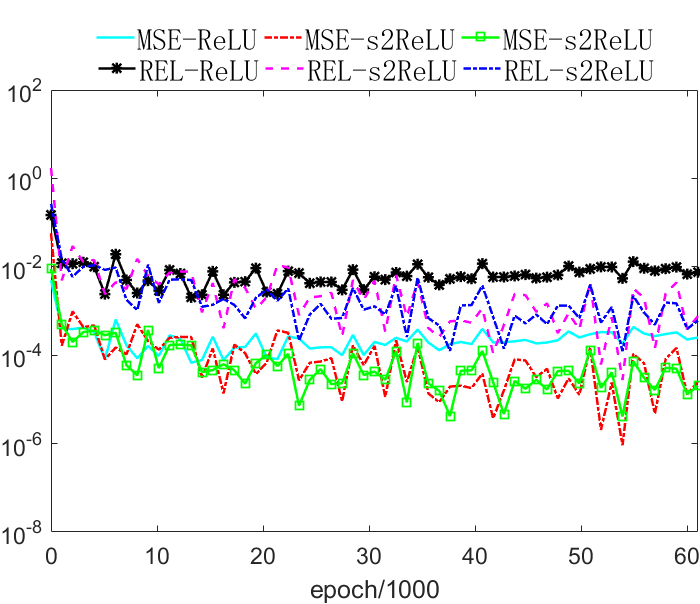}
	}
	\caption{Testing results for $\epsilon=0.01$ when $p=5$. The network size is (500, 400, 300, 300, 200, 100, 100).}
	\label{p5e001}
\end{figure}

We show the testing results for $\epsilon=0.1$ and $\epsilon=0.01$  in Figs. \ref{p5e01} and \ref{p5e001}, respectively. 
The MscaleDNN with activation function s2ReLU can well capture all the oscillation of the exact solution for $\epsilon=0.1$ in Fig. \ref{p5e01a}, which is much better than that of sReLU and ReLU, and the test error of s2ReLU is much lower as shown in Fig. \ref{p5e01b}. For $\epsilon=0.01$, MscaleDNNs still outperform activation function ReLU, while s2ReLU and sReLU are comparable, as shown in Fig. \ref{p5e001}. 

From the above results, we conclude that the MscaleDNN model with s2ReLU activation function can much better solve the p-Laplacian problem compared with the ones of sReLU and ReLU, even for a nonlinear case.

\subsection{Two dimensional examples}
We consider the following p-Laplacian problem in domain $\Omega=[-1,1]\times[-1,1]$,
\begin{equation}\label{Eq0402}
\begin{cases}
-\text{div}\bigg{(}\kappa(x_1,x_2)|\nabla u(x_1,x_2)|^{p-2} \nabla u(x,y)\bigg{)} = f(x_1,x_2),\\
u(-1,x_2)=u(1,x_2)=0, u(x_1, -1)=u(x_1, 1)=0.
\end{cases}
\end{equation}

In the following tests, we obtain the solution of \eqref{Eq0402} by employing two types of MscaleDNN with size (1000, 500, 400, 300, 300, 200, 100, 100) and activation functions sReLU and s2ReLU, respectively. 
Based on the conclusions of MscaleDNN for one-dimensional p-Laplacian problems and previous results for MscaleDNN in solving PDEs \cite{liu2020multi}, a MscaleDNN  with s2ReLU or sReLU  outperforms DNN with ReLU, therefore, we will not show the results of DNN  with ReLU in the following experiments.

\begin{example}\label{2dE02}
	In this example, the forcing term $f(x_1,x_2)\equiv 1$ for $p=2$ and a multi-scale trigonometric coefficient $\kappa(x_1,x_2)$ is given by
	\begin{align*}
	\kappa(x_1,x_2) = &\frac{1}{6}\left(\frac{1.1+\sin(2\pi x_1/\epsilon_1)}{1.1+\sin(2\pi x_2/\epsilon_1)}+\frac{1.1+\sin(2\pi x_1/\epsilon_2)}{1.1+\cos(2\pi x_2/\epsilon_2)}+\frac{1.1+\cos(2\pi x_1/\epsilon_3)}{1.1+\sin(2\pi x_2/\epsilon_3)}\right.\\
	&\left.+\frac{1.1+\sin(2\pi x_1/\epsilon_4)}{1.1+\cos(2\pi x_2/\epsilon_4)} + \frac{1.1+\cos(2\pi x_1/\epsilon_5)}{1.1+\sin(2\pi x_2/\epsilon_5)}+\sin(4x_1^2x_2^2)+1
	\right),
	\end{align*}
	where $\displaystyle \epsilon_1=\frac{1}{5},\epsilon_2=\frac{1}{13},\epsilon_3=\frac{1}{17},\epsilon_4=\frac{1}{31},\epsilon_5=\frac{1}{65}.$ For this example, the corresponding exact solution can not be expressed explicitly. Alternatively, a reference solution $u(x_1,x_2)$ is set as the finite element solution computed by  numerical homogenization method \cite{owhadi2007homogenization,owhadi2008numerical,owhadi2014polyharmonic}  on a square grid $[-1,1]\times[-1,1]$ of mesh-size $h=(1+2^q)^{-1}$ with a positive integer $q=6$. 
\end{example}

At each training step, we randomly sample $n_{it} = 3000$ interior points and $n_{bd}=500$ boundary points as training dataset. The testing dataset are also sampled from a square grid $[-1,1]\times[-1,1]$ of mesh-size $h=(1+2^q)^{-1}$ with $q=6$. 

\begin{figure} [H]
	\centering    
 	\subfigure[Cut lines of solutions] {
 		\label{2dPDE2q6:a}     
 		\includegraphics[scale=0.30]{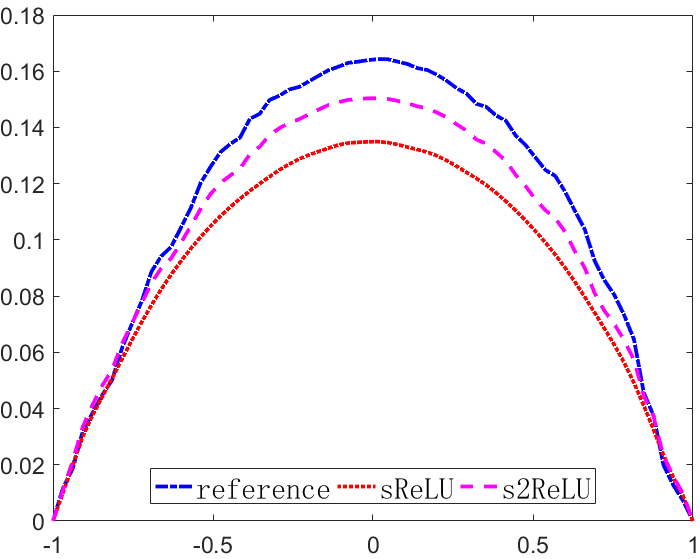}  
 	}    
	\subfigure[MSE and REL] { 
		\label{2dPDE2q6:f}     
		\includegraphics[scale=0.395]{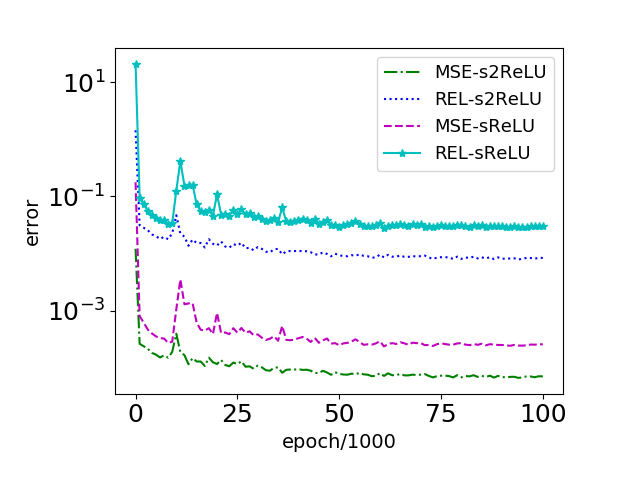}     
	}
	\subfigure[point-wise error] {
		\label{2dPDE2q6:d}     
		\includegraphics[scale=0.275]{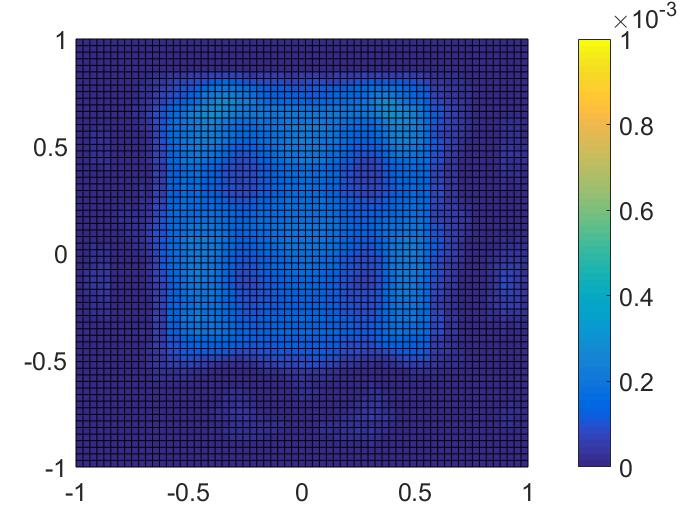}  
	}     
	\subfigure[point-wise error] { 
		\label{2dPDE2q6:e}     
		\includegraphics[scale=0.275]{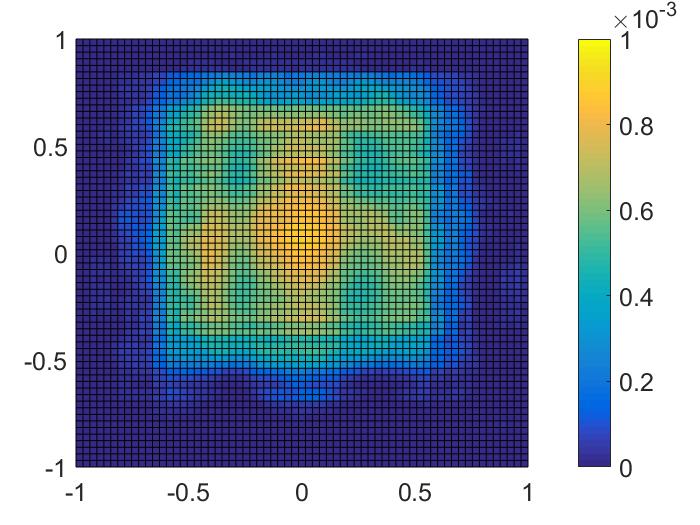}     
	}    
	
	\caption{Testing results for Example \ref{2dE02}. {\ref{2dPDE2q6:a}: Cut lines along $x=0$ for reference solution, s2ReLU solution and sReLU solution, respectively.}  \ref{2dPDE2q6:f}: Mean square error and relative error for s2ReLU and sReLU, respectively. \ref{2dPDE2q6:d}: Point-wise square error for s2ReLU. \ref{2dPDE2q6:e}: Point-wise square error for sReLU. } 
	\label{2dPDE2q6}         
\end{figure}

As shown in Figs. ~\ref{2dPDE2q6:a} and \ref{2dPDE2q6:f}, for the high-frequency oscillatory coefficient $\kappa(x_1,x_2)$ in this example, the performances of our model with s2ReLU and sReLU are still favorable to solve \eqref{Eq0402} and our s2ReLU performs better than sReLU in overall training process. Figs. \ref{2dPDE2q6:d} and \ref{2dPDE2q6:e} not only show that the point-wise errors for major points are closed to zero, but also reveal that the point-wise error of s2ReLU is smaller than that of sReLU.  In short, our model with s2ReLU activation function  can obtain a satisfactory solution for p-Laplacian problem and it outperforms the one of sReLU.

\begin{example}\label{2dE05}
	In this example, we test the performance of MscaleDNN to p-Laplacian problem for $p=3$. The forcing term $f(x_1,x_2)$ and $\kappa(x_1,x_2)$ are similar to that in Example \ref{2dE02}. Analogously, we still take the reference solution $u$ as the finite element solution on a fine mesh over the square domain $[0,1]\times[0,1]$ of mesh-size $h=(1+2^q)^{-1}$ with a positive integer $q=6$. In addition, the training and testing datasets in this example are similarly constructed as the Example \ref{2dE02}.
\end{example}

\begin{figure} [H]
	\centering    
    \subfigure[Cut lines of solutions] { 
		\label{2dPDE5q6:a}     
		\includegraphics[scale=0.30]{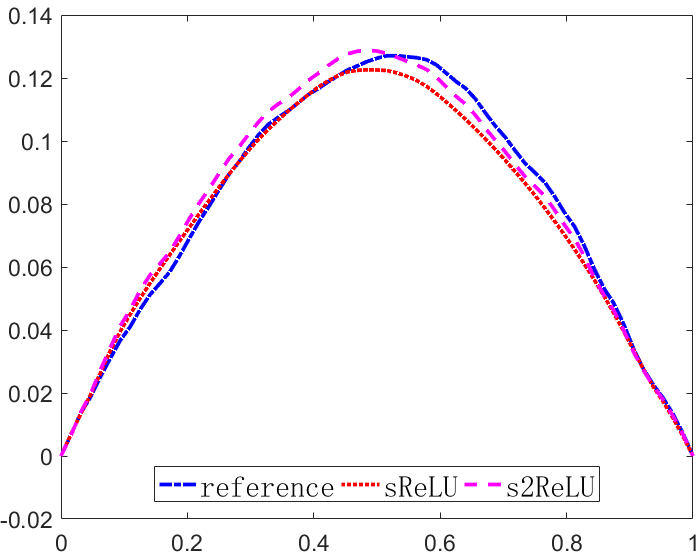}     
	}
	\subfigure[MSE and REL] { 
		\label{2dPDE5q6:f}     
		\includegraphics[scale=0.395]{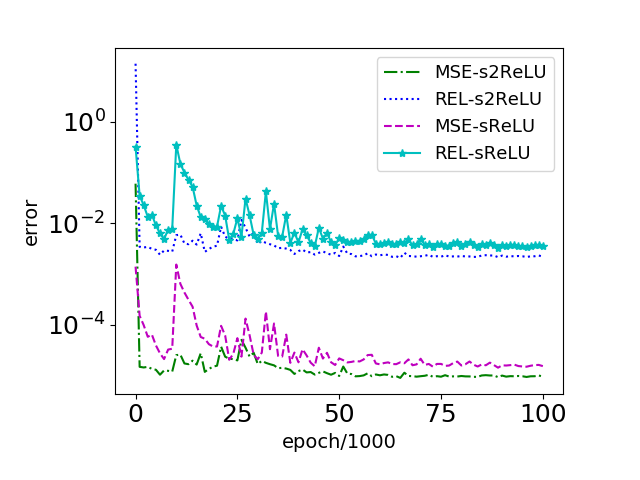}     
	}
	\subfigure[point-wise error] {
		\label{2dPDE5q6:d}     
		\includegraphics[scale=0.275]{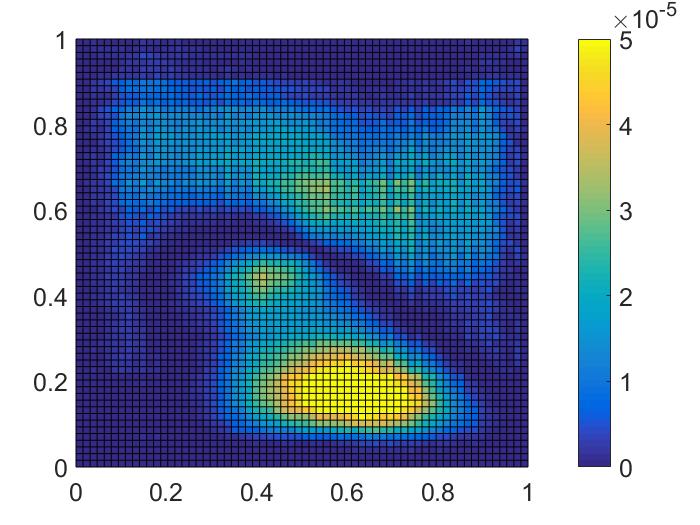}  
	}     
	\subfigure[point-wise error] { 
		\label{2dPDE5q6:e}     
		\includegraphics[scale=0.275]{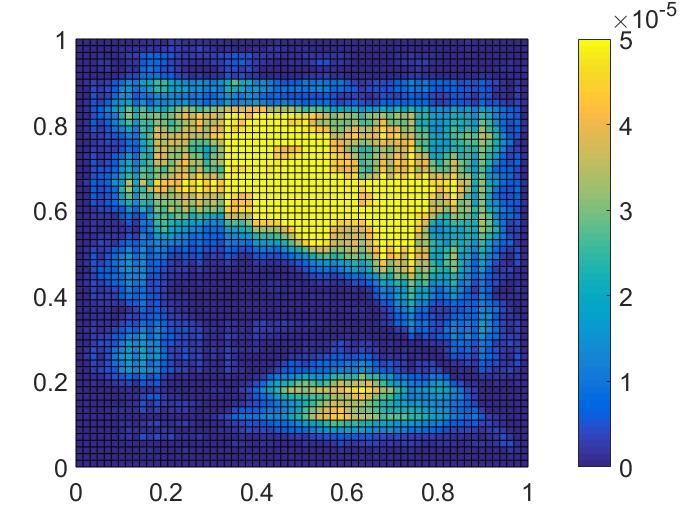}     
	}        
\caption{Testing results for Example \ref{2dE05}. {\ref{2dPDE5q6:a}: Cut lines along $x=0.5$ for reference solution, s2ReLU solution and sReLU solution, respectively.} \ref{2dPDE5q6:f}: Mean square error and relative error for s2ReLU and sReLU, respectively. \ref{2dPDE5q6:d}: Point-wise square error for s2ReLU. \ref{2dPDE5q6:e}: Point-wise square error for sReLU.} 
	\label{2dPDE5q6}         
\end{figure}

From the results in Figure \ref{2dPDE5q6}, the performance of MscaleDNN with s2ReLU is also superior to the one of sReLU. The overall errors (including MSE and REL) of both activation functions are comparable, but the point-wise error of s2ReLu is  smaller than that of sReLU .

\begin{example}\label{2dE03}
	In this example, we take the forcing term $f=1$ for  $p=2$,  and 
	\begin{equation*}
	\kappa(x_1,x_2) = \Pi_{k=1}^q\bigg{(}1+0.5\cos\left(2^k\pi(x_1+x_2)\right)\bigg{)}\bigg{(}1+0.5\sin\left(2^k\pi(x_2-3x_1)\right)\bigg{)}
	\end{equation*}
	where $q$ is a positive integer. The coefficient $\kappa(x_1,x_2)$ has non-separable scales. Similarly to Example \ref{2dE02}, we take the reference solution $u$ as the finite element solution on a fine mesh over the square domain $[-1,1]\times[-1,1]$ of mesh-size $h=(1+2^q)^{-1}$ with a positive integer $q=6$. 
\end{example}

\begin{figure} [H]
	\centering    
	\subfigure[Cut lines of solutions] { 
		\label{2dPDE3q6:a}     
		\includegraphics[scale=0.30]{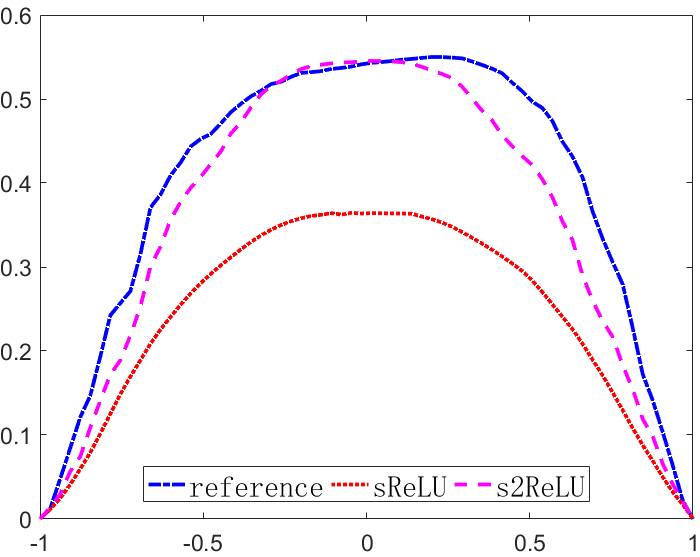}     
	}
	\subfigure[MSE and REL] { 
		\label{2dPDE3q6:f}     
		\includegraphics[scale=0.395]{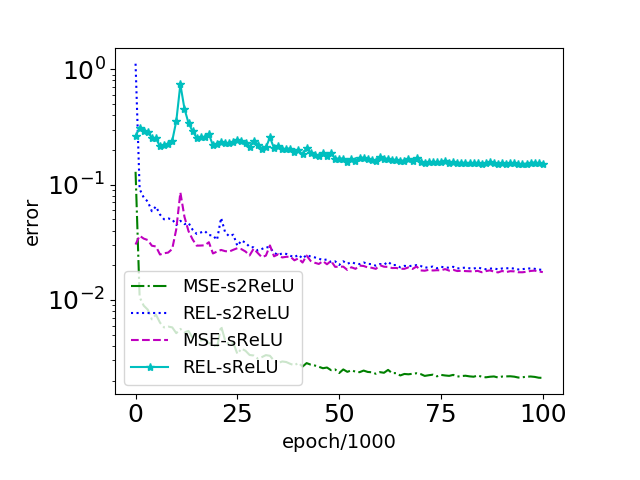}     
	}
	\subfigure[point-wise error] {
		\label{2dPDE3q6:d}     
		\includegraphics[scale=0.275]{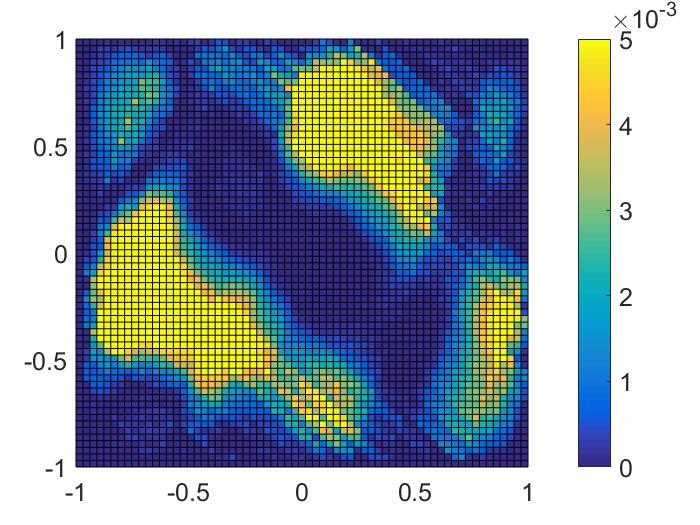}  
	}     
	\subfigure[point-wise error] { 
		\label{2dPDE3q6:e}     
		\includegraphics[scale=0.275]{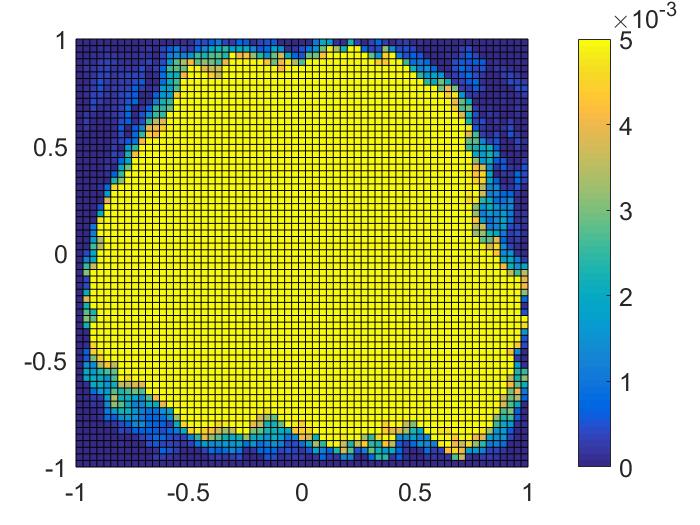}     
	}    
	\caption{Testing results for Example \ref{2dE03}. {\ref{2dPDE3q6:a} : Cut lines along $x=0$ for reference solution, s2ReLU solution and sReLU solution, respectively.} \ref{2dPDE3q6:f}: Mean square error and relative error for s2ReLU and sReLU, respectively. \ref{2dPDE3q6:d}: Point-wise square error for s2ReLU. \ref{2dPDE3q6:e}: Point-wise square error for sReLU. }   
	\label{2dPDE3q6}       
\end{figure}

In this example, the training and testing datasets are similarly constructed as the Example \ref{2dE02}.  In  Figs. \ref{2dPDE3q6:a} and \ref{2dPDE3q6:f}, the s2ReLU solution approximates the exact solution much better than that of sReLU solution. This can be clearly seen from the point-wise error in Figs. \ref{2dPDE3q6:d} and \ref{2dPDE3q6:e}. 

Based on the results of the two dimensional examples \ref{2dE02}, \ref{2dE05} and \ref{2dE03}, it is clear that the MscaleDNN model with s2ReLU activation function can approximate the solution of mulitscale elliptic problems with oscillating coefficients with possible nonlinearity, and its performance is superior to that of sReLU. It is important to examine the capability of MscaleDNN for high-dimensional (multi-scale) elliptic problems, which will be shown in the following.

\subsection{High dimensional examples}

\begin{example}\label{E5d}
	We consider the following p-Laplacian problem in  domain $\Omega=[0,1]^{5}$
	\begin{equation}\label{Eq04011}
	\begin{cases}
	-\text{div}\bigg{(}\kappa(x_1,x_2,\cdots,x_{5})|\nabla u(x_1,x_2,\cdots,x_{5})|^{p-2} \nabla u(x_1,x_2,\cdots,x_{5})\bigg{)} = f(x_1,x_2,\cdots,x_{5}),\\
	u(0,x_2,\cdots,x_{5})=u(1,x_2,\cdots,x_{5})=0, \\
	~~~~~~~~~~~~~~~~~~~\cdots\cdots\\
	u(x_1,x_2,\cdots,0)=u(x_1,x_2,\cdots,1)=0.
	\end{cases}
	\end{equation}
	In this example, we take $p=2$ and 
	\begin{equation*}
	\kappa(x_1,x_2,\cdots,x_{5})=1+\cos(\pi x_1)\cos(2\pi x_2)\cos(3\pi x_3)\cos(2\pi x_4)\cos(\pi x_5).
	\end{equation*}
	We choose the forcing term $f$ such that the exact solution is 
	\begin{equation*}
	u(x_1,x_2,\cdots,x_{5})=\sin(\pi x_1)\sin(\pi x_2)\sin(\pi x_{3})\sin(\pi x_{4})\sin(\pi x_{5}).
	\end{equation*}
\end{example}

\begin{figure} [H]
	\centering 
	\subfigure[MSE and REL] { 
		\label{5dPDE1:f}     
		\includegraphics[scale=0.285]{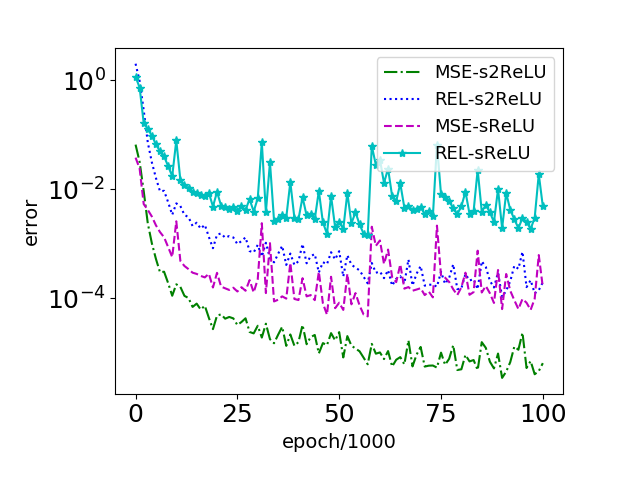}     
	}
	\subfigure[point-wise error] {
		\label{5dPDE1:d}     
		\includegraphics[scale=0.19]{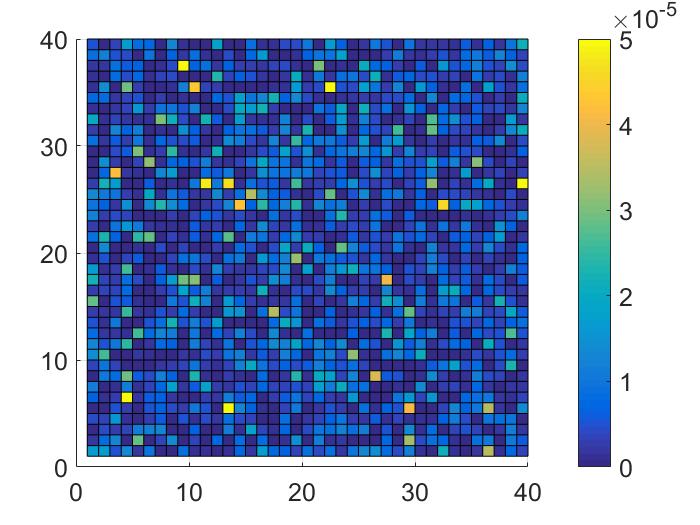}  
	}     
	\subfigure[point-wise error] { 
		\label{5dPDE1:e}     
		\includegraphics[scale=0.19]{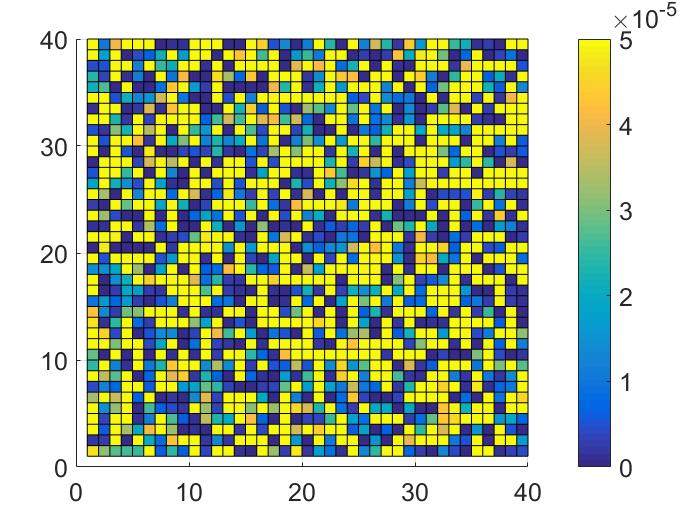}     
	}    
	\caption{Testing results for Example \ref{E5d}. \ref{5dPDE1:f}: Mean square error and relative error for s2ReLU and sReLU, respectively. \ref{5dPDE1:d}: Point-wise square error for s2ReLU. \ref{5dPDE1:e}: Point-wise square error for sReLU. } 
	\label{5dPDE1}         
\end{figure}

For five-dimensional elliptic problems, we use two types of MscaleDNNs with size (1000, 800, 500, 500, 400, 200, 200, 100) and activation functions s2ReLU and sReLU, respectively. 
The training data set includes 7500 interior points and 1000 boundary points randomly sampled from $\Omega$. The testing dataset includes 1600 random samples in $\Omega$. We plot the testing results in Fig. \ref{5dPDE1}. To visually illustrate these results, we map the point-wise errors of sReLU and s2ReLU solutions, evaluated on 1600 sample points in $\Omega$, onto a $40\times 40$ 2d array, respectively. We note that the mapping is only for the purpose of visualization, and is independent of the actual coordinates of those points.

The numerical results in Fig. \ref{5dPDE1:f} indicate that the MscaleDNN models with s2ReLU and sReLU can still well approximate the exact solution of elliptic equation in five-dimensional space. In particular, Figs. \ref{5dPDE1:d} and \ref{5dPDE1:e} show that the point-wise error of s2ReLU is much smaller than that of sReLU. 

\section{Conclusion}\label{sec:05}
In this paper, we propose an improved version of MscaleDNN by designing an activation function localized in both spatial and Fourier domains, and use that to solve multi-scale elliptic problems. Numerical results show that this method is effective for the resolution of elliptic problems with multiple scales and possible nonlinearity, in low to median high dimensions. As a meshless method, DNN based method is more flexible for partial differential equations than traditional mesh-based and meshfree methods in regular or irregular region. In the future, we will optimize the MscaleDNN architecture and design DNN based algorithms for multi-scale nonlinear problems with more general nonlinearities.

\section*{Acknowledgements}
X.L and L.Z  are partially supported by the National Natural Science Foundation of China (NSFC 11871339, 11861131004). Z.X. is supported by National Key R\&D Program of China (2019YFA0709503), and Shanghai Sailing Program. This work is also partially supported by HPC of School of Mathematical Sciences at Shanghai Jiao Tong University.


\end{document}